\documentclass{llncs} \usepackage{makeidx} \usepackage{subfigure}

\usepackage{times} \usepackage{epsfig,wrapfig}
\usepackage[english]{babel}
\usepackage{url}  \usepackage{ifsym}
\usepackage{amsmath,amssymb,latexsym}
\usepackage{multirow}

\input xy \xyoption{all}

\catcode`\@=11\relax
\newwrite\@unused
\def\typeout#1{{\let\protect\string\immediate\write\@unused{#1}}}
\def\figurepath{./}

\def\@nnil{\@nil}
\def\@empty{}
\def\@psdonoop#1\@@#2#3{}
\def\@psdo#1:=#2\do#3{\edef\@psdotmp{#2}\ifx\@psdotmp\@empty \else
    \expandafter\@psdoloop#2,\@nil,\@nil\@@#1{#3}\fi}
\def\@psdoloop#1,#2,#3\@@#4#5{\def#4{#1}\ifx #4\@nnil \else
       #5\def#4{#2}\ifx #4\@nnil \else#5\@ipsdoloop #3\@@#4{#5}\fi\fi}
\def\@ipsdoloop#1,#2\@@#3#4{\def#3{#1}\ifx #3\@nnil 
       \let\@nextwhile=\@psdonoop \else
      #4\relax\let\@nextwhile=\@ipsdoloop\fi\@nextwhile#2\@@#3{#4}}
\def\@tpsdo#1:=#2\do#3{\xdef\@psdotmp{#2}\ifx\@psdotmp\@empty \else
    \@tpsdoloop#2\@nil\@nil\@@#1{#3}\fi}
\def\@tpsdoloop#1#2\@@#3#4{\def#3{#1}\ifx #3\@nnil 
       \let\@nextwhile=\@psdonoop \else
      #4\relax\let\@nextwhile=\@tpsdoloop\fi\@nextwhile#2\@@#3{#4}}
\def\psdraft{
	\def\@psdraft{0}
}
\def\psfull{
	\def\@psdraft{100}
}
\psfull
\newif\if@prologfile
\newif\if@postlogfile
\newif\if@noisy
\def\pssilent{
	\@noisyfalse
}
\def\psnoisy{
	\@noisytrue
}
\psnoisy
\newif\if@bbllx
\newif\if@bblly
\newif\if@bburx
\newif\if@bbury
\newif\if@height
\newif\if@width
\newif\if@scale
\newif\if@rheight
\newif\if@rwidth
\newif\if@clip
\newif\if@verbose
\def\@p@@sclip#1{\@cliptrue}

\def\@p@@sfile#1{\def\@p@sfile{null}%
	        \openin1=#1
		\ifeof1\closein1%
		       \openin1=\figurepath#1
			\ifeof1\typeout{Error, File #1 not found}
			\else\closein1
			    \edef\@p@sfile{\figurepath#1}%
                        \fi%
		 \else\closein1%
		       \def\@p@sfile{#1}%
		 \fi}
\def\@p@@sfigure#1{\def\@p@sfile{null}%
	        \openin1=#1
		\ifeof1\closein1%
		       \openin1=\figurepath#1
			\ifeof1\typeout{Error, File #1 not found}
			\else\closein1
			    \def\@p@sfile{\figurepath#1}%
                        \fi%
		 \else\closein1%
		       \def\@p@sfile{#1}%
		 \fi}

\def\@p@@sbbllx#1{
		\@bbllxtrue
		\dimen100=#1
		\edef\@p@sbbllx{\number\dimen100}
}
\def\@p@@sbblly#1{
		\@bbllytrue
		\dimen100=#1
		\edef\@p@sbblly{\number\dimen100}
}
\def\@p@@sbburx#1{
		\@bburxtrue
		\dimen100=#1
		\edef\@p@sbburx{\number\dimen100}
}
\def\@p@@sbbury#1{
		\@bburytrue
		\dimen100=#1
		\edef\@p@sbbury{\number\dimen100}
}
\def\@p@@sscale#1{
		\@scaletrue
		\count255=#1
   		\edef\@p@sscale{\number\count255}
}
\def\@p@@sheight#1{
		\@heighttrue
		\dimen100=#1
   		\edef\@p@sheight{\number\dimen100}
}
\def\@p@@swidth#1{
		\@widthtrue
		\dimen100=#1
		\edef\@p@swidth{\number\dimen100}
}
\def\@p@@srheight#1{
		\@rheighttrue
		\dimen100=#1
		\edef\@p@srheight{\number\dimen100}
}
\def\@p@@srwidth#1{
		\@rwidthtrue
		\dimen100=#1
		\edef\@p@srwidth{\number\dimen100}
}
\def\@p@@ssilent#1{ 
		\@verbosefalse
}
\def\@p@@sprolog#1{\@prologfiletrue\def\@prologfileval{#1}}
\def\@p@@spostlog#1{\@postlogfiletrue\def\@postlogfileval{#1}}
\def\@cs@name#1{\csname #1\endcsname}
\def\@setparms#1=#2,{\@cs@name{@p@@s#1}{#2}}
\def\ps@init@parms{
		\@bbllxfalse \@bbllyfalse
		\@bburxfalse \@bburyfalse
		\@heightfalse \@widthfalse
		\@scalefalse
		\@rheightfalse \@rwidthfalse
		\def\@p@sbbllx{}\def\@p@sbblly{}
		\def\@p@sbburx{}\def\@p@sbbury{}
		\def\@p@sheight{}\def\@p@swidth{}
		\def\@p@sscale{}
		\def\@p@srheight{}\def\@p@srwidth{}
		\def\@p@sfile{}
		\def\@p@scost{10}
		\def\@sc{}
		\@prologfilefalse
		\@postlogfilefalse
		\@clipfalse
		\if@noisy
			\@verbosetrue
		\else
			\@verbosefalse
		\fi
}
\def\parse@ps@parms#1{
	 	\@psdo\@psfiga:=#1\do
		   {\expandafter\@setparms\@psfiga,}}
\newif\ifno@bb
\newif\ifnot@eof
\newread\ps@stream
\def\bb@missing{
	\if@verbose{
		\typeout{psfig: searching \@p@sfile \space  for bounding box}
	}\fi
	\openin\ps@stream=\@p@sfile
	\no@bbtrue
	\not@eoftrue
	\catcode`\%=12
	\loop
		\read\ps@stream to \line@in
		\global\toks200=\expandafter{\line@in}
		\ifeof\ps@stream \not@eoffalse \fi
		\@bbtest{\toks200}
		\if@bbmatch\not@eoffalse\expandafter\bb@cull\the\toks200\fi
	\ifnot@eof \repeat
	\catcode`\%=14
}	
\catcode`\%=12
\newif\if@bbmatch
\def\@bbtest#1{\expandafter\@a@\the#1
\long\def\@a@#1
\long\def\bb@cull#1 #2 #3 #4 #5 {
	\dimen100=#2 bp\edef\@p@sbbllx{\number\dimen100}
	\dimen100=#3 bp\edef\@p@sbblly{\number\dimen100}
	\dimen100=#4 bp\edef\@p@sbburx{\number\dimen100}
	\dimen100=#5 bp\edef\@p@sbbury{\number\dimen100}
	\no@bbfalse
}
\catcode`\%=14
\def\compute@bb{
		\no@bbfalse
		\if@bbllx \else \no@bbtrue \fi
		\if@bblly \else \no@bbtrue \fi
		\if@bburx \else \no@bbtrue \fi
		\if@bbury \else \no@bbtrue \fi
		\ifno@bb \bb@missing \fi
		\ifno@bb \typeout{FATAL ERROR: no bb supplied or found}
			\no-bb-error
		\fi
		\count203=\@p@sbburx
		\count204=\@p@sbbury
		\advance\count203 by -\@p@sbbllx
		\advance\count204 by -\@p@sbblly
		\edef\@bbw{\number\count203}
		\edef\@bbh{\number\count204}
}
\def\in@hundreds#1#2#3{\count240=#2 \count241=#3
		     \count100=\count240	
		     \divide\count100 by \count241
		     \count101=\count100
		     \multiply\count101 by \count241
		     \advance\count240 by -\count101
		     \multiply\count240 by 10
		     \count101=\count240	
		     \divide\count101 by \count241
		     \count102=\count101
		     \multiply\count102 by \count241
		     \advance\count240 by -\count102
		     \multiply\count240 by 10
		     \count102=\count240	
		     \divide\count102 by \count241
		     \count200=#1\count205=0
		     \count201=\count200
			\multiply\count201 by \count100
		 	\advance\count205 by \count201
		     \count201=\count200
			\divide\count201 by 10
			\multiply\count201 by \count101
			\advance\count205 by \count201
		     \count201=\count200
			\divide\count201 by 100
			\multiply\count201 by \count102
			\advance\count205 by \count201
		     \edef\@result{\number\count205}
}
\def\compute@wfromh{
		\in@hundreds{\@p@sheight}{\@bbw}{\@bbh}
		\edef\@p@swidth{\@result}
}
\def\compute@hfromw{
		\in@hundreds{\@p@swidth}{\@bbh}{\@bbw}
		\edef\@p@sheight{\@result}
}
\def\compute@wfroms{
		\in@hundreds{\@p@sscale}{\@bbw}{100}
		\edef\@p@swidth{\@result}
}
\def\compute@hfroms{
		\in@hundreds{\@p@sscale}{\@bbh}{100}
		\edef\@p@sheight{\@result}
}
\def\compute@handw{
		\if@scale
			\compute@wfroms
			\compute@hfroms
		\else
			\if@height 
				\if@width
				\else
					\compute@wfromh
				\fi	
			\else 
				\if@width
					\compute@hfromw
				\else
					\edef\@p@sheight{\@bbh}
					\edef\@p@swidth{\@bbw}
				\fi
			\fi
		\fi
}
\def\compute@resv{
		\if@rheight \else \edef\@p@srheight{\@p@sheight} \fi
		\if@rwidth \else \edef\@p@srwidth{\@p@swidth} \fi
}
\def\compute@sizes{
	\compute@bb
	\compute@handw
	\compute@resv
}
\def\psfig#1{\vbox {
	\ps@init@parms
	\parse@ps@parms{#1}
	\compute@sizes
	\ifnum\@p@scost<\@psdraft{
		\if@verbose{
			\typeout{psfig: including \@p@sfile \space }
		}\fi
		\special{ps::[begin] 	\@p@swidth \space \@p@sheight \space
				\@p@sbbllx \space \@p@sbblly \space
				\@p@sbburx \space \@p@sbbury \space
				startTexFig \space }
		\if@clip{
			\if@verbose{
				\typeout{(clip)}
			}\fi
			\special{ps:: doclip \space }
		}\fi
		\if@prologfile
		    \special{ps: plotfile \@prologfileval \space } \fi
		\special{ps: plotfile \@p@sfile \space }
		\if@postlogfile
		    \special{ps: plotfile \@postlogfileval \space } \fi
		\special{ps::[end] endTexFig \space }
		\vbox to \@p@srheight true sp{
			\hbox to \@p@srwidth true sp{
				\hss
			}
		\vss
		}
	}\else{
		\vbox to \@p@srheight true sp{
		\vss
			\hbox to \@p@srwidth true sp{
				\hss
				\if@verbose{
					\@p@sfile
				}\fi
				\hss
			}
		\vss
		}
	}\fi
}}
\def\psglobal{\typeout{psfig: PSGLOBAL is OBSOLETE; use psprint -m instead}}
\catcode`\@=12\relax

\pagestyle{empty}

\begin{document}

\newcommand{\node}[1]{{\tt#1}}
\newcommand{\figeps}[3]{ 
  \begin{figure}[h!tb]
    \def\epsfsize##1##2{#2##1}%
    \centerline{\epsfbox{Graphics/#1.eps}}
    \caption{#3}
    \label{fig:#1}
  \end{figure}
}

\newcommand{\proofend}{\hfill$\blacksquare$}
\newcommand{\trace}{\mathrm{tr}}

\title{Matrix Graph Grammars as a Model of Computation}
 \author{Pedro Pablo P\'erez Velasco}

\institute{
  Escuela Polit\'ecnica Superior\\
  Universidad Aut\'onoma de Madrid\\
  \email{pedro.perez@uam.es} }

\maketitle

\begin{abstract}
  Matrix Graph Grammars (MGG) is a novel approach to the study of
  graph dynamics (\cite{MGG_Book}). In the present contribution we
  look at MGG as a formal grammar and as a model of computation, which
  is a necessary step in the more ambitious program of tackling
  complexity theory through MGG. We also study its relation with other
  well-known models such as Turing machines (TM) and Boolean circuits
  (BC) as well as non-determinism. As a side effect, all techniques
  available for MGG can be applied to TMs and BCs.
\end{abstract}

\textbf{Keywords}: Matrix Graph Grammars, Graph Dynamics, Graph
Transformation, Graph Rewriting, Model of Computation.

\section{Introduction}
\label{sec:intro}

Graph transformation~\cite{graGraBook,handbook} is becoming
increasingly popular in order to describe system behavior due to its
graphical, declarative and formal nature. It is central to many
application areas, such as visual languages, visual simulation,
picture processing and model transformation (see~\cite{graGraBook} and
\cite{handbook} Vol.2 for some applications). Also, graph rewriting
techniques have proved useful in describing \emph{Domain Specific
  Languages}\footnote{Following~\cite{DSL}, a domain-specific
  language (DSL) is a programming language or executable specification
  language that offers, through appropriate notations and
  abstractions, expressive power focused on, and usually restricted
  to, a particular problem domain.} and in language-oriented
programming (see~\cite{JVLC}).

Matrix Graph Grammars (MGG) is a purely algebraic approach to graph
rewriting (graph dynamics) that has successfully solved or extended
problems and results such as sequentialization, explicit parallelism,
applicability, congruence characterization, initial state calculation
(initial digraphs), constraints (application conditions) and
reachability. See for
example~\cite{JuanPP_1,MGG_ICGT,MGG_PNGT,MGG_PROLE,MGG_Combinatorics,MGG_Fundamenta}
or the more comprehensive introduction~\cite{MGG_Book}.

It seems natural to study MGG as a model of computation that would
describe the state of some system by means of graphs. To the very best
of the author's knowledge, there has been no attempt in this direction
until now, despite the agreed interest of the topic (see
e.g.~\cite{GCM1,GCM2}). Even more so, there seems to be no graph
rewriting literature on non-determinism or, more generally, on
complexity.

There are two main handicaps that may partially explain the lack of
theoretical results. First, graph rewriting systems (MGG in
particular) would make use of the functional problem known as
\emph{subgraph matching}. Its associated decision problem
(\emph{subgraph isomorphism}) has been proved to be
\textbf{NP}-complete~(\cite{Garey}). Even worse, subgraph matching
would be used in every single step of the computation. Second, there
are two apparently hard-to-avoid sources of non-determinism: what
production rule to apply and where to apply it.

The author's main motivation for the development of MGG is the study
of complexity theory -- complexity classes \textbf{P} and \textbf{NP}
in particular -- which has led the research in MGG up to
now.\footnote{Almost all concepts and problems addressed
  in~\cite{MGG_Book} directly study sequentialization. Recall that
  \textbf{P}, \textbf{NP} and in general the classes in $\mathcal{PH}$
  (the polynomial hierarchy) encode sequentialization.} The present
contribution is a necessary step in this program.

Due to their current theoretical and practical relevance, we have
decided to study Turing machines (TM) and Boolean Circuits (BC) and
compare them to MGG. As a side effect, all techniques developed for
MGG become available to study TMs and BCs. Currently we are capable to
(partially) answer the following questions within MGG:
\begin{enumerate}
\item Coherence: are the actions specified by the productions (inside
  a sequence) \emph{compatible} with each other?
\item Initial state: calculate an initial configuration (initial
  grammar state) such that some given sequence can be applied.
\item Congruence: do a sequence and a permutation of it have a common
  initial state?
\item Sequential independence: is it possible to advance/delay some
  production a finite number of positions inside a given sequence? 
\item Reachability: for a given grammar and initial and final states,
  find a sequence\footnote{$\ldots$ or provide information about it,
    e.g. what production rules have to be applied and how many times
    each production should be applied.} that transforms the initial
  state into the final state.
\end{enumerate}

\noindent \textbf{Paper Organization}. Section~\ref{sec:MGGbasics} is
an overview of the very basics of MGG and briefly explains some of the
analysis techniques developed so far. Section~\ref{sec:relabeling}
characterizes relabeling in
MGG. Section~\ref{sec:formalGrammar}~and~\ref{sec:modelOfComputation}
study MGG as a formal grammar and as a model of computation,
respectively, touching on some possible submodels and supermodels of
computation that can be defined from MGG. Determinism is addressed in
Secs.~\ref{sec:nonDeterminism}~and~\ref{sec:BA2VS}, in which we move
from Boolean algebra to an algebra of matrices. In
Secs.~\ref{sec:MGGandTM}~and~\ref{sec:MGGandBC} MGG models Turing
machines and Boolean Circuits,
respectively. Section~\ref{sec:conclusions} closes this paper with a
short summary and some proposals for future research.

\noindent \textbf{Notation}. The matrix whose entries are all zero
will be represented with a bolded zero, \textbf{0}. Similarly, the
matrix in which every single element is a one will be represented with
a bolded one, \textbf{1}.

\section{Matrix Graph Grammars: Basics}
\label{sec:MGGbasics}

In this section we give a brief overview of some of the basics of
Matrix Graph Grammars (MGG) with examples as intuitive as possible.
For a detailed account and accessible presentation the reader is
referred to~\cite{MGG_Book}.
\\ \
\\ \
\noindent \textbf{Simple Digraphs}. We work with simple digraphs,
which we represent as $(M, V)$ where $M$ is a Boolean matrix for edges
(the graph {\em adjacency} matrix) and $V$ a Boolean vector for
vertices or nodes. Note that we explicitly represent the nodes of the
graph with a vector. This is necessary because it is possible within
MGG to add and delete nodes.

\begin{figure}[htbp]
  \centering \subfigure{
    \includegraphics[scale = 0.52]{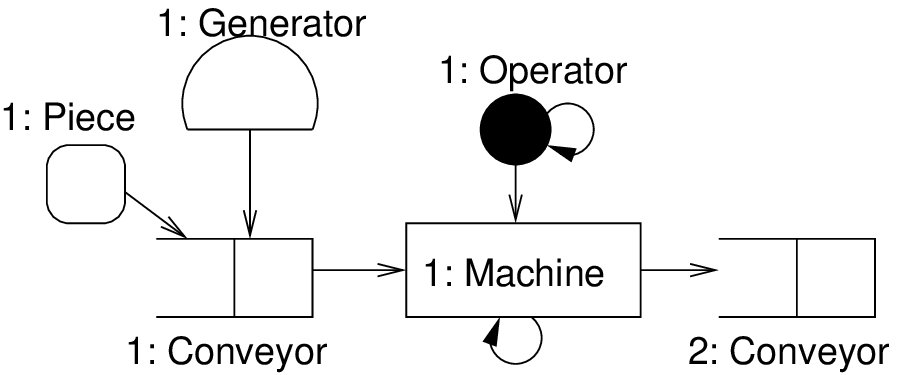}}
  \hspace{0.3cm} \subfigure{
    \includegraphics[scale = 0.65]{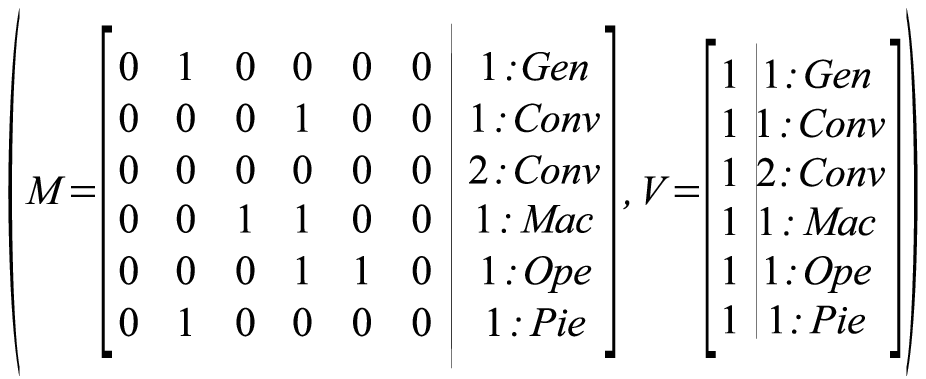}}
  \caption{Simple Digraph Example (left). Matrix Representation with
    Labels (right)}
  \label{fig:example_graph}
\end{figure}

The existing nodes are marked with a $1$ in the corresponding position
of the vector. Figure~\ref{fig:example_graph}(a) shows a graph
representing a production system made of a machine (controlled by an
operator) which consumes and produces pieces through
conveyors. Generators create pieces in conveyors. Self loops in
operators and machines indicate that they are busy.

Note that the matrix and the vector in the figure are the smallest
ones able to represent the graph. Adding $0$-elements to the vector
(and accordingly $0$-rows and columns to the matrix) would result in
equivalent graphs. Next definition formulates the representation of
simple digraphs.

\begin{definition}[Simple Digraph Representation]
  \label{def:simple_digraph}
  A simple digraph $G$ is represented by $G=(M, V)$ where $M$ is the
  graph adjacency matrix and $V$ is the Boolean vector of its nodes.
\end{definition}

\noindent \textbf{Compatibility}. Well-formedness of graphs (i.e.
absence of dangling edges) can be checked by verifying the identity
$\left\| \left( M \vee M^t \right) \odot \overline{V}\right\| _1 = 0$,
where $\odot$ is the Boolean matrix product (like the regular matrix
product, but with {\bf and} and {\bf or} instead of multiplication and
addition), $M^t$ is the transpose of the matrix $M$, $\overline{V}$ is
the negation of the nodes vector $V$, and $\| \cdot \|_1$ is an
operation (a norm, actually) that results in the {\bf or} of all the
components of the vector. This property is called \emph{compatibility}
in~\cite{JuanPP_1}. Note that $M \odot \overline{V}$ results in a
vector that contains a $1$ in position $i$ when there is an outgoing
edge from node $i$ to a non-existing node. A similar expression with
the transpose of $M$ is used to check for incoming edges. A simple
digraph $G = (M, V)$ is compatible if and only if
\begin{equation}
  \label{eq:11}
  \left\| \left( M \vee M^t \right) \odot \overline{V}\right\|_1 = 0.
\end{equation}

Compatibility of productions guarantees that the image of a simple
digraph is a simple digraph again. It is useful to check closedness of
the space (simple digraphs) under the specified operations (grammar
rules).
\\ \
\\ \
\noindent \textbf{Labeling}. A label (also known as a \emph{type}) is
assigned to each node in $G = (M,V)$ by a function from the set of
nodes $V$ to a set of labels $\mathcal{T}$, $\lambda: V \rightarrow
\mathcal{T}$. In Fig.~\ref{fig:example_graph} labels are represented
as an extra column in the matrices, the numbers before the colon
distinguish elements of the same type.

There are several equivalent possibilities to label edges. We may use
the types of their source and target nodes. Another possibility is to
choose the set of edges $E(G)$ as domain for $\lambda$ (see
Def.~\ref{def:typed_simple_digraph}) instead of $V$. Notice that this
would define $\lambda$ for every element in the adjacency
matrix.\footnote{Similarly, labels on the nodes would be equivalent to
  defining $\lambda$ for the elements in the main diagonal of the
  adjacency matrix.} Yet another possibility is to define labels just
on nodes and split one edge $N_1N_2$ (the one joining node $N_1$ to
node $N_2$) into two edges, the first starting in node $N_1$ and
ending in a ``labeling node'' and the second starting in this same
``labeling node'' and ending in $N_2$.

\begin{definition}[Labeled Simple Digraph]
  \label{def:typed_simple_digraph}
  A labeled simple digraph $G_\lambda = (G, \lambda) = ((M, V),
  \lambda)$ over a set of labels $\mathcal{T}$ is made of a simple
  digraph $G$ plus a function $\lambda: V \rightarrow \mathcal{T}$
  whose domain is the set of nodes of $G$.
\end{definition}

By abuse of notation, the subscript $\lambda$ will normally be
omitted. Next we define the notion of partial morphism between labeled
simple digraphs.

\begin{definition}[Labeled Simple Digraph Morphism]
  \label{def:morphism}
  Given two simple digraphs $G_i=((M_i, V_i), \lambda_i: V_i
  \rightarrow \mathcal{T})$ for $i=\{1, 2\}$, a morphism $f=(f^V,
  f^E):G_1 \rightarrow G_2$ is made of two partial injective
functions $f^V:V_1 \rightarrow V_2$, $f^E: M_1 \rightarrow M_2$
between the set of nodes and edges, such that:
\begin{enumerate}
\item $\forall v \in Dom(f^V), \: \lambda_1(v) = \lambda_2 \left(
    f^V(v) \right)$
\item $\forall (n, m) \in Dom(f^E), \: f^E(n, m) = \left( f^V(n),
  f^V(m) \right)$,
\end{enumerate}
where $Dom(f)$ is the domain of the partial function $f$, $E$ stands
for edges and $V$ for vertices.
\end{definition}

\noindent \textbf{Productions}. A production, $p:L \rightarrow R$ is a
morphism of labeled simple digraphs. Using a {\em static formulation},
a rule is represented by two labeled simple digraphs that encode the
left and right hand sides (LHS and RHS, respectively). The matrices
and vectors of these graphs are arranged so that the elements
identified by morphism $p$ match (this is called completion, see
below).

\begin{definition}[Static Formulation of Productions]
  \label{def:static_production}
  A production $p:L \rightarrow R$ is statically represented as
  \begin{equation}
    \label{eq:13}
    p = ( L, R ) = \left( (L^E, L^V, \lambda_L), (R^E, R^V, \lambda_R)
    \right).
  \end{equation}
\end{definition}

A production adds and deletes nodes and edges, therefore using a {\em
  dynamic formulation}, we can encode the rule's pre-condition (its
LHS) together with matrices and vectors representing the addition and
deletion of edges and nodes. We call such matrices and vectors $e$ for
``erase'' and $r$ for ``restock''.

\begin{definition}[Dynamic Formulation of Productions]
  \label{def:dynamic_production}
  A production $p:L \rightarrow R$ is dynamically represented as
  \begin{equation}
    \label{eq:14}
    p = ( L, e, r) = \left( \left( L^E, L^V, \lambda_L \right) ,
      \left( e^E, e^V, \lambda_e \right) , \left( r^E, r^V, \lambda_r
      \right) \right),
  \end{equation}
  where $e^E$ and $e^V$ are the deletion Boolean matrix and vector
  (respectively), $r^E$ and $r^V$ are the addition Boolean matrix and
  vector (respectively). They have a $1$ in the position where the
  element is to be deleted (for $e$) or added (for $r$).
\end{definition}

The output of rule $p$ -- where the \textbf{and} symbol $\wedge$ is
omitted to simplify formulae\footnote{In order to avoid ambiguity we
  shall state that $\wedge$ has precedence over $\vee$.} -- is
calculated by the Boolean formula
\begin{equation}
  \label{eq:15}
   R = p(L) = r \vee \overline{e} L,
\end{equation}
which applies to both, edges and nodes.

\begin{figure}[htbp]
  \centering
  \subfigure{
    \includegraphics[scale = 0.4]{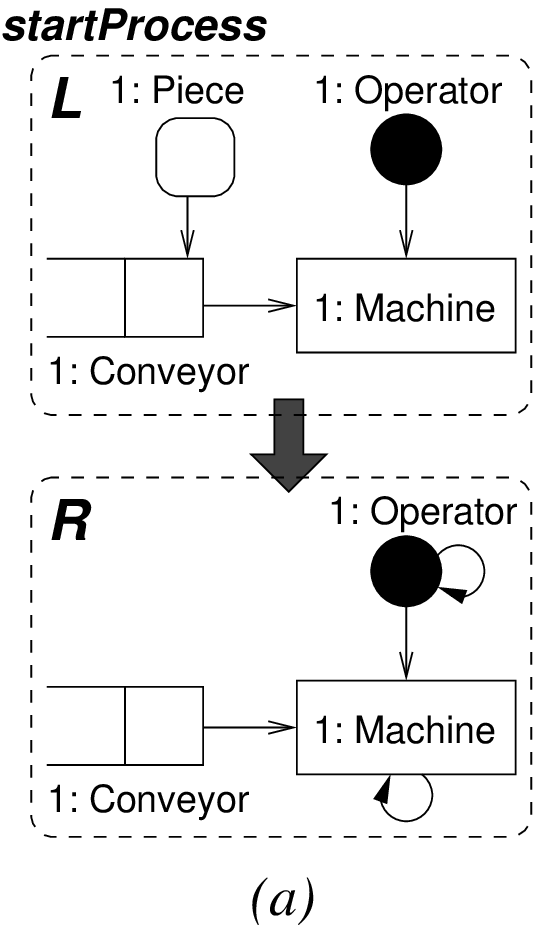}}
  \hspace{0.4cm}
  \subfigure{
    \includegraphics[scale = 0.5]{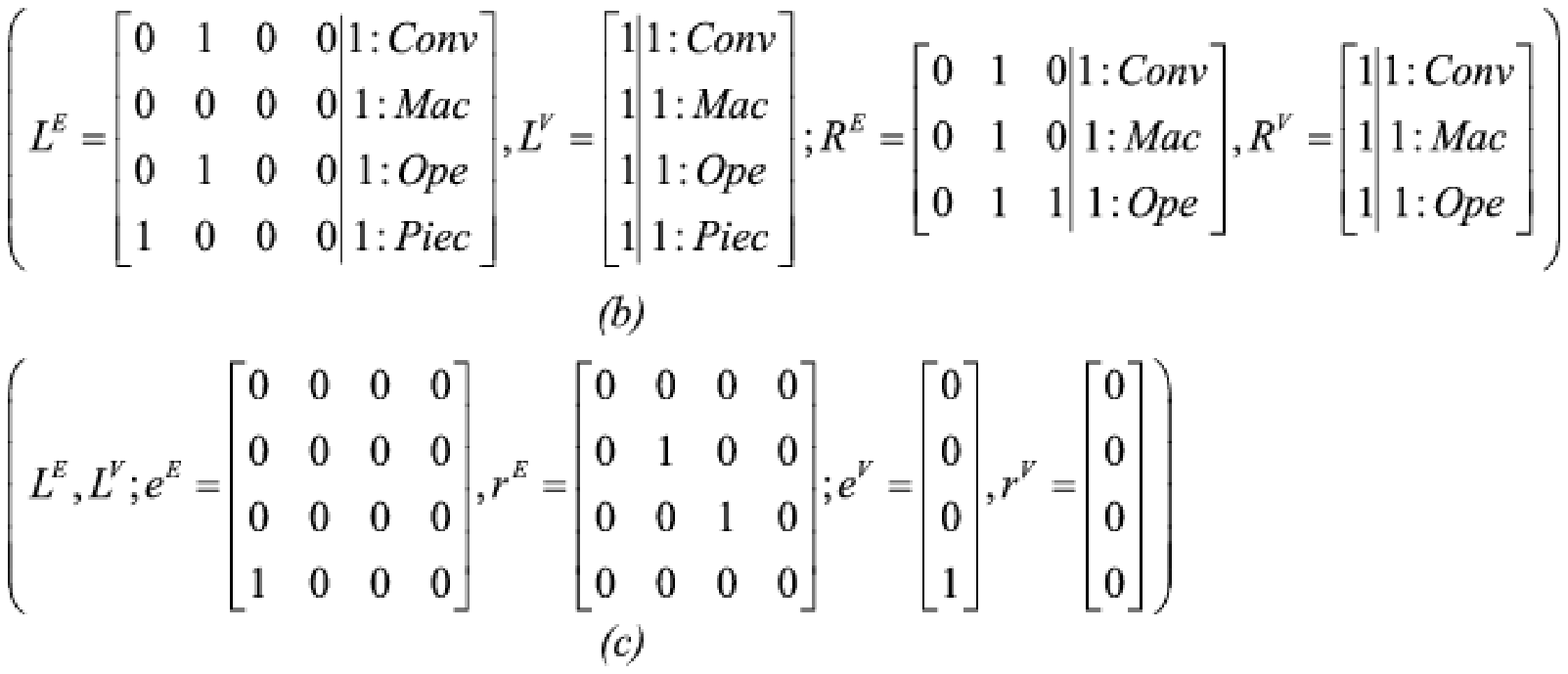}}
  \caption{(a) Rule Example. (b) Static Formulation. (c) Dynamic
    Formulation.}
  \label{fig:example_rule}
\end{figure}

\noindent \textbf{Example}.$\square$Figure~\ref{fig:example_rule}
shows a rule and its associated matrices. The rule models the
consumption of a piece by a machine.  Compatibility of the resulting
graph must be ensured, thus the rule cannot be applied if the machine
is already busy, as it would end up with two self loops, which is not
allowed in a simple digraph.  This restriction of simple digraphs can
be useful in this kind of situations, and acts like a built-in
negative \emph{application condition} (refer to~\cite{MGG_Book},
Chaps.~8~and~9). Later we will see that the \emph{nihilation matrix}
takes care of this restriction.

\noindent \textbf{Completion}. In order to operate graphs of different
sizes, an operation called completion adds extra rows and columns with
zeros (to matrices and vectors) and rearranges rows and columns so
that the identified edges and nodes of the two graphs match.  For
example, in Fig.~\ref{fig:example_rule}, if $L^E$ and $R^E$ need to be
operated, completion adds a fourth 0-row and fourth 0-column to $R^E$.

Otherwise stated, whenever we have to operate graphs $G_1$ and
$G_2$, an implicit morphism $f:G_1 \rightarrow G_2$ has to be
established. This morphism is completion, which rearranges the
matrices and the vectors of both graphs so that the elements match. In
the examples, we omit such operation, and assume that matrices are
completed when necessary.
\\ \
\\ \
\noindent \textbf{Nihilation Matrix}. In order to consider the
elements in the host graph that disable a rule application, we extend
the notation for rules with a new graph $K$. Its associated matrix
specifies the two kinds of forbidden edges: those incident to nodes
which are going to be erased and any edge added by the rule (which
cannot be added twice, since we are dealing with simple
digraphs). Notice however that $K$ considers only potential dangling
edges with source and target in the nodes belonging to $L^V$.

\begin{definition}[Nihilation Matrix]
  \label{def:nihilation_matrix}
  Given the production $p = ( L, e, r)$, its nihilation matrix $K^E$
  contains non-zero elements in positions corresponding to newly added
  edges, and to non-deleted edges adjacent to deleted nodes.
\end{definition}

We extend the rule formulation with this nihilation matrix. The
concept of production remains unaltered because we are just making
explicit some implicit information. Matrices are derived in the
following order: $\left( L, R \right) \mapsto \left( e, r \right)
\mapsto K$.  Thus, a rule is \emph{statically} determined by its LHS
and RHS $p = \left( L, R \right)$, from which it is possible to give a
dynamic definition $p = \left(L, e, r \right)$, with $e =
L\overline{R}$ and $r = R\overline{L}$, to end up with a full
specification including its \emph{environmental} behavior $p =
\left(L, K, e, r \right)$.  No extra effort is needed from the grammar
designer because $K$ can be automatically calculated as the image by
the rule $p$ of a certain matrix (see below).

Notice that the nihilation matrix $K$ for a production $p$ has an
associated simple digraph whose nodes coincide with those of $L$ as
well as its labels.


\begin{definition}[Dynamic Reformulation of Production]
  \label{def:dynamic_reformulation_production}
  A production $p:L \rightarrow R$ is dynamically represented as
  \begin{align}
    \label{eq:16}
    p &= ( L, K, e, r, \lambda) = \nonumber \\
    &= \left( \left( L^E, L^V, \lambda_L \right), \left( K^E, K^V,
        \lambda_K \right), \left( e^E, e^V, \lambda_e \right), \left(
        r^E, r^V, \lambda_r \right) \right),
  \end{align}
  where $L$ is the LHS, $K$ is the labeled simple digraph associated
  to the nihilation matrix, $e$ is the deletion matrix and $r$ is the
  addition matrix.
\end{definition}

The nihilation matrix $K^E$ of a given production $p$ is given by
\begin{equation}
  \label{eq:41}
  K^E = p \left(\overline{D} \right) \quad \mathrm{ with } \quad D =
  \overline{e^V} \otimes \overline{e^V}^t,
\end{equation}
where $\otimes$ denotes the tensor (or Kronecker) product, which sums
up the covariant and contravariant parts and multiplies every element
of the first vector by the whole second
vector. See~\cite{MGG_Fundamenta} for a proof. Notice that matrix
$\overline{D}$ specifies potential dangling edges incident to nodes in
$p$'s LHS:
\begin{equation*}
  \overline{D} = d^i_{\!j} = \left\{
    \begin{array}{ll}
      1 & \qquad if \; (e^V)^i = 1 \; or \; (e^V)^j = 1. \\
      0 & \qquad otherwise.
    \end{array} \right.
\end{equation*}

\noindent \textbf{Example}.$\square$The nihilation matrix $K^E$ for
the example rule of Fig.~\ref{fig:example_rule} is calculated as
follows:

\begin{equation}
  \overline{\overline{e^V} \otimes \overline{\left( e^V \right)}^{\,
      t}} = 
  \overline{ \left[
      \begin{array}{c}
        1 \\ 1 \\ 1 \\ 0 \\
      \end{array}
    \right] \otimes
    \left[
      \begin{array}{c}
        1 \\ 1 \\ 1 \\ 0 \\
      \end{array}
    \right]^t} =
  \left[
    \begin{array}{cccc}
      0 & 0 & 0 & 1 \\
      0 & 0 & 0 & 1 \\
      0 & 0 & 0 & 1 \\
      1 & 1 & 1 & 1 \\
    \end{array}
  \right]
  \nonumber
\end{equation}

The nihilation matrix is then given by :
\begin{equation}
  K = r \vee \overline e \overline{D} =
  \left[
    \begin{array}{cccc}
      0 & 0 & 0 & 0\\
      0 & 1 & 0 & 0\\
      0 & 0 & 1 & 0\\
      0 & 0 & 0 & 0\\
    \end{array}
  \right] \vee
  \overline {\left[
      \begin{array}{cccc}
        0 & 0 & 0 & 0\\
        0 & 0 & 0 & 0\\
        0 & 0 & 0 & 0\\
        1 & 0 & 0 & 0 \\
      \end{array} 
    \right]}
  \left[
    \begin{array}{cccc}
      0 & 0 & 0 & 1 \\
      0 & 0 & 0 & 1 \\
      0 & 0 & 0 & 1 \\
      1 & 1 & 1 & 1 \\
    \end{array} 
  \right] =
  \left[
    \begin{array}{cccc}
      0 & 0 & 0 & 1 \\
      0 & 1 & 0 & 1 \\
      0 & 0 & 1 & 1 \\
      0 & 1 & 1 & 1 \\
    \end{array}
  \right] \nonumber
\end{equation}

The matrix indicates any dangling edge from the deleted piece (the
edge to the conveyor is not signaled as it is explicitly deleted) as
well as self-loops in the machine and in the operator. \proofend


\begin{figure}
  \centering
  \includegraphics[width=0.25\textwidth]{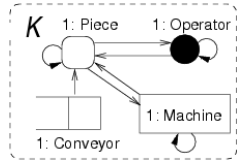}
  \caption{$K$ Graph for \emph{startProcess}}
  \label{fig:nihilation_example}
\end{figure}

The evolution of the rule LHS (i.e. how it is transformed into the
RHS) is given by the production itself: $R = p(L) = r \vee
\overline{e} \, L$. See equation~(\ref{eq:15}). It is interesting to
analyze the behavior of the nihilation matrix, which is given in the
next proposition. Let $p:L \rightarrow R$ be a compatible production
with nihilation graph $K$. Then, the elements that must not appear
($Q$) once the production is applied (see~\cite{MGG_Fundamenta} for a
proof) are given by
\begin{equation}
  \label{eq:42}
  Q = p^{-1} \left( K \right) = e \vee \overline{r} \, K^E.
\end{equation}


\begin{figure}[htbp]
  \centering
  \includegraphics[scale = 0.45]{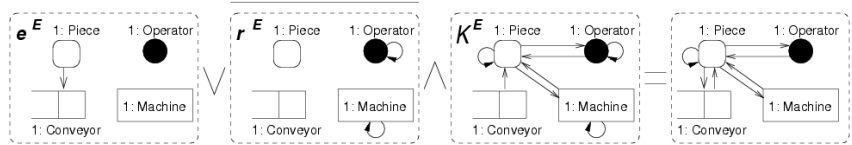}
  \caption{Evolution of the Nihilation Matrix}
  \label{fig:nihilation_evol}
\end{figure}

\noindent \textbf{Example}.$\square$Figure~\ref{fig:nihilation_evol}
shows the calculation of $startProcess^{-1}(K^E)$ using the graph
representation of the matrices in eq.~(\ref{eq:42}). \proofend

The dual concept $T$ specifies the newly available edges after the
application of a production due to the addition of nodes:
\begin{equation}
  \label{eq:43}
  T = \left( \overline{ \overline{r} \otimes \overline{r}^t} \right)
  \wedge \left( \overline{e} \otimes \overline{e}^t \right).
\end{equation}
Refer to~\cite{MGG_Combinatorics} where $T$ was introduced and
studied.

Next definition introduces a functional notation for rules (already
used in~\cite{MGG_ICGT}) inspired by the Dirac or bra-ket
notation~\cite{braket}.

\begin{definition}[Functional Formulation of Production]
  \label{def:functional_rule}
  A production $p:L \rightarrow R$ can be depicted as $R = p(L) =
  \left\langle L, p \right\rangle$, splitting the static part (initial
  state, $L$) from the dynamics (element addition and deletion, $p$).
\end{definition}

Using such formulation, the \emph{ket} operators (i.e. those to the
right of the bra-ket) can be moved to the \emph{bra} (i.e. left hand
side) by using their adjoints (which are usually decorated with an
asterisk).
\\ \
\\ \
\noindent \textbf{Match and Derivations}. Matching is the operation of
identifying the LHS of a rule inside a host graph (we consider only
injective matches). Given the rule $p:L \rightarrow R$ and a simple
digraph $G$, any $m:L \rightarrow G$ total injective morphism is a
match for $p$ in $G$. The following definition considers not only the
elements that should be present in the host graph $G$ (those in $L$)
but also those that should not be (those in $K$).

\begin{definition}[Direct Derivation]
  \label{def:directDerivationDef}
  Given the rule $p:L \rightarrow R$ and the graph $G=(G^E, G^V)$ as
  in Fig.~\ref{fig:matches}(a), $d = \left( p, m \right)$ -- with $m =
  \left( m_L, m^E_K \right)$ -- is called a \emph{direct derivation}
  with result $H = p^* \left( G \right)$ if the following conditions
  are fulfilled:
  \begin{enumerate}
  \item There exist $m_L : L \rightarrow G$ and $m^E_K: K^E
    \rightarrow \overline{G^E}$ total injective morphisms.
  \item $m_L(n) = m^E_N(n)$, $\forall n \in L^V$.
  \item The match $m_L$ induces a completion of $L$ in $G$. Matrices
    $e$ and $r$ are then completed in the same way to yield $e^*$ and
    $r^*$. The output is given by $H = p^*(G) = r^* \vee
    \overline {e^*} G$.
  \end{enumerate}

\begin{figure}[htbp]
  \centering
  \includegraphics[scale = 0.42]{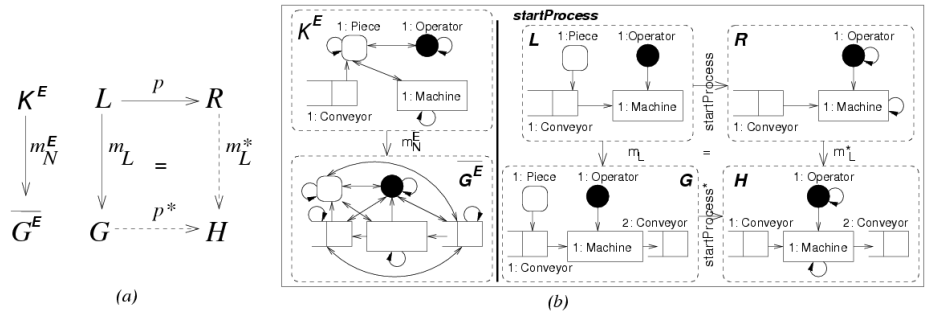}
  \caption{(a) Direct Derivation. (b) Example}
  \label{fig:matches}
\end{figure}

\end{definition}

\noindent \textbf{Remark}.$\square$The square in
Fig.~\ref{fig:matches} (a) is a categorical pushout. Item 2 is needed
to ensure that $L$ and $K^E$ are matched to the same nodes in
$G$. \proofend

\noindent \textbf{Example}.$\square$Figure~\ref{fig:matches}(b) shows
the application of rule \emph{startProcess} to graph $G$. We have also
depicted the inclusion of $K^E$ in $\overline {G^E}$ (bidirectional
arrows have been used for simplification). $\overline {G^E}$ is the
complement (negation) of matrix $G^E$. \proofend

It is useful to consider the structure defined by the negation of the
host graph, $\overline G=(\overline {G^E}, \overline{G^V})$.  It is
made up of the graph $\overline {G^E}$ and the vector of nodes
$\overline{G^V}$. Note that the negation of a graph is not a graph
because in general compatibility fails, that is why the term
``structure'' is used.


The complement of a graph coincides with the negation of the adjacency
matrix, but while negation is just the logical operation, taking the
complement means that a completion has been performed in advance. That
is, the complement of graph $G$ with respect to graph $A$, through a
morphism $f:A \rightarrow G$ is a two-step operation: (i) complete $G$
and $A$ according to $f$, yielding $G'$ and $A'$; (ii) negate $G'$.
As long as no confusion arises negation and complements will not be
distinguished syntactically.

\begin{figure}[htbp]
  \centering
  \includegraphics[scale = 0.45]{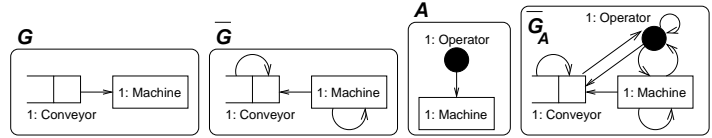}
  \caption{Finding Complement and Negation of a Graph.}
  \label{fig:negation_graph}
\end{figure}

\noindent \textbf{Examples}.$\square$Suppose we have two graphs $A$
and $G$ as those depicted in Fig.~\ref{fig:negation_graph} and that we
want to check that $A$ is not in $G$.  Note that $A$ is not contained
in $\overline G$ (an operator node does not even appear) but it does
appear in the negation of the completion of $G$ with respect to $A$
(graph $\overline{G}_{A}$ in the same figure).

In the context of Fig.~\ref{fig:matches}(b) we see that there is an
inclusion $startProcess^{-1}(K^E) \rightarrow \overline H$ (i.e. the
forbidden elements after applying production $startProcess$ are not in
$H$). This is so because we complete $H$ with an additional piece
(which was deleted from $G$). \proofend
\\ \
\\ \
\noindent \textbf{Analysis Techniques.}
In~\cite{JuanPP_1,MGG_ICGT,MGG_PNGT,MGG_PROLE,MGG_Book} we developed
some analysis techniques, mainly to study sequences in MGG. We end
this section with a short summary of these techniques with the
exception of application conditions, graph constraints and explicit
parallelism.

One of the goals of our previous work was to analyze rule sequences
independently of a host graph. We represent a rule sequence as
$s_n=p_n; ...; p_1$, where application is from right to left (i.e.
$p_1$ is applied first). For its analysis, we complete the sequence by
identifying the nodes across rules which are assumed to be mapped to
the same node in the host graph. Mind the non-commutativity of this
operation and its potential non-determinism.

Once the sequence is completed, our notion of sequence
\emph{coherence}~\cite{JuanPP_1} allows to know if, for the given
identification, the sequence is potentially applicable (i.e. if no
rule disturbs the application of those following it). The formula for
coherence results in a matrix and a vector (which can be interpreted
as a graph) with the problematic elements. If the sequence is
coherent, both should be zero, if not, they contain the problematic
elements. We shall elaborate on this in
Secs.~\ref{sec:nonDeterminism}~and~\ref{sec:BA2VS}.

A coherent sequence is {\em compatible} if its application
produces a simple digraph. That is, no dangling edges are produced in
intermediate steps. This extends the notions of compatible graph and
compatible production.

Given a completed sequence, the minimal initial digraph (MID) is the
smallest graph that allows applying such sequence. Conversely, the
negative initial digraph (NID) contains all elements that should not
be present in the host graph for the sequence to be applicable. In
this way, the NID is a graph that should be found in $\overline G$ for
the sequence to be applicable (i.e. none of its edges can be found in
$G$). We shall elaborate on this in Sec.~\ref{sec:BA2VS}.

We shall not touch on other concepts such as G-congruence and
reachability. Notice that reachability can be thought of as an initial
attempt to measure the number of elements in a sequence, a tool that
might be useful to provide lower bounds. Similarly to what is done in
the present contribution for Turing Machines and Boolean Circuits,
Petri nets can be modeled with Matrix Graph
Grammars. See~\cite{MGG_Combinatorics}~and~\cite{MGG_Book}.



\section{Relabeling}
\label{sec:relabeling}

Completion as introduced in Sec.~\ref{sec:MGGbasics} performs two
tasks: enlarges matrices by adding rows and columns of the appropriate
type and rearranges them (to get a coincidence according to the
identifications across productions). The first task -- as we shall see
in Sec.~\ref{sec:formalGrammar} -- has to do with the dimension of the
underlying algebraic structure. The second is directly related to
non-determinism, for which we need to give an \emph{operational}
definition. The section is dedicated to this topic.

Relabeling is just a permutation on nodes. A simple observation is
that to any permutation $\sigma \in S_n$, a permutation Al matrix
$\sigma$ can be associated with $\sigma^i_j = 1 \Leftrightarrow
\sigma(i) = j$. Notice that a Boolean matrix is a permutation matrix
with respect to the action defined in eq.~(\ref{eq:17}) if and only if
it has a single $1$ per row and column. Its action $\rho$ is defined
by
\begin{equation}
  \label{eq:17}
  \rho_\sigma(L) = \sigma(L) = \sigma . \, L = \sigma \odot L \odot
  \sigma^t,
\end{equation}
where $\odot$ is the matrix product but with \textbf{or} operations
instead of sums and with \textbf{and} instead of multiplication, and
$L$ is the adjacency matrix of some simple digraph. For a quick
introduction on permutation matrices, please refer to
\cite{Weisstein2}. We shall also use $\cdot$ in place of $\odot$.

As a matter of fact, eq.~(\ref{eq:17}) defines a production as it
transforms one simple digraph into another simple digraph.\footnote{By
  abuse of notation we shall represent the permutation, its associated
  matrix and the relabeling production that it defines with the same
  letter $\sigma$.} As such, it can be expressed in terms of some
appropriate $e$ and $r$ matrices:
\begin{equation}
  \label{eq:18}
  e = L \left( \,\overline{\sigma.\,L} \,\right) = L \left( \,
    \overline{\sigma \cdot L \cdot \sigma^t} \, \right), \qquad r =
  \overline{L} (\sigma.\,L) = \overline{L} \left( \sigma \cdot L \cdot
    \sigma^t \right).
\end{equation}
Recall that an elementwise \textbf{and} is assumed when the operation
is omitted. If $A$ and $B$ are $n \times n$ Boolean matrices then $AB
= \left( a^i_j \wedge b^i_j \right)_{i,j \in \{1, \ldots, n\}}$.

\begin{figure}[htbp]
  \centering
  \includegraphics[scale = 0.6]{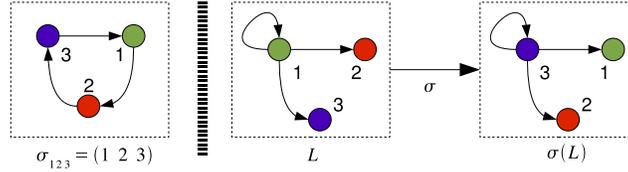}
  \caption{Example of Permutation (Left) and its Associated
    Production (Right)}
  \label{fig:permRepresentation}
\end{figure}

\noindent \textbf{Example}.$\square$Suppose we are given the
permutation $\sigma = (1\; 2\; 3) \in S_3$ to be applied to the graph
$L$, depicted to the right of Fig.~\ref{fig:permRepresentation}. The
associated permutation matrix is
\begin{displaymath}
  \sigma = \left[\begin{array}{ccc}
      \vspace{-4pt}
      0 & 1 & 0 \\
      \vspace{-4pt}
      0 & 0 & 1 \\
      \vspace{-4pt}
      1 & 0 & 0 \\
      \vspace{-10pt}
    \end{array} \right].
\end{displaymath}
Its action is equivalent to a node relabeling where node $1$ plays the
role of node $2$, node $2$ that of node $3$ and node $3$ becomes node
$1$. If we wanted to put $\sigma$ as an $(e,r)$ production, its erasing
and addition matrices would be:
\begin{displaymath}
  e_\sigma = \left[\begin{array}{ccc}
      \vspace{-4pt}
      1 & 1 & 1 \\
      \vspace{-4pt}
      0 & 0 & 0 \\
      \vspace{-4pt}
      0 & 0 & 0 \\
      \vspace{-10pt}
    \end{array} \right] \qquad
  r_\sigma = \left[\begin{array}{ccc}
      \vspace{-4pt}
      0 & 0 & 0 \\
      \vspace{-4pt}
      0 & 0 & 0 \\
      \vspace{-4pt}
      1 & 1 & 1 \\
      \vspace{-10pt}
    \end{array} \right],
\end{displaymath}
which have been calculated using eq.~(\ref{eq:18}).

Say we need to relabel node $3$ in Fig. \ref{fig:permRepresentation}
with a new type, for example $4$. It would be possible to proceed in
three stages: first add node $4$, then permute nodes $3$ and $4$
applying the 2-cycle $\sigma_{34} = (3 \; 4)$ -- its graph will have
nodes $1$ and $2$ with self edges -- and then delete node
$3$. \proofend

Despite the possibility of reducing $\odot$ to some $(e,r)$
productions, we shall introduce this new operation, extending the
notion of production to that of \emph{affine production}. The main
reason is that, according to eq.~(\ref{eq:18}), the relabeling
expressed in terms of $(e,r)$ (as a production) would have different
matrices depending on the graph it is applied to. The action of
relabeling should have associated a single element, independently of
the simple digraph it acts on.

Through concatenation we obtain the two possible combinations of
operations $p;\sigma$ and $\sigma;p$, which are respectively:
\begin{align}
  \label{eq:21}
  p;\sigma(L) &=  r \vee \overline{e} \left( \sigma \cdot L \cdot
    \sigma^t \right) \\
  \label{eq:22}
  \sigma;p(L) &=  \sigma \cdot \left( r \vee \overline{e} L \right)
  \cdot \sigma^t
\end{align}
We shall be more interested in the combination of both
$\sigma;p;\sigma$ -- permute the graph, apply production and permute
the result. It is readily seen that
\begin{align}
  \label{eq:31}
  p^\sigma(L) = \sigma;p;\sigma(L) &= \sigma \cdot \left[ r \vee
    \overline{e} \left( \sigma \cdot L \cdot \sigma^t \right) \right]
  \cdot \sigma^t = \left[ \sigma \cdot \left( r \vee \overline{e}
    \right) \cdot \sigma^t \right] L,
\end{align}
which defines the action $\rho_\sigma$ of eq.~(\ref{eq:17}) for
productions. Equation~\eqref{eq:31} can be rewritten $p^\sigma(L) =
\langle L, \sigma.\,p \rangle$.


\begin{definition}[Affine Production]
  \label{def:full_dynamic_production}
  A production $p:L \rightarrow R$ as defined by eq.~(\ref{eq:31}) is
  dynamically represented by
  \begin{equation}
    \label{eq:23}
    p = ( L, K, e, r, \sigma, \lambda)
  \end{equation}
  where $L$ is the LHS, $K$ is the labeled simple digraph associated
  to the nihilation matrix, $e$ is the deletion matrix, $r$ is the
  addition matrix, $\sigma$ is a relabeling and $\lambda$ is the node
  labeling mapping.
\end{definition}

Clearly, making $\sigma$ the identity returns an $(e,r)$ production
(which will be known as the \emph{traslational part} of the affine
production) and setting $e = r = \mathbf{0}$ transforms the production
in a \emph{pure} relabeling action (which will be known as the
\emph{rotational part} of the affine production).

\begin{proposition}
  \label{prop:homomorphism}
  The action on productions $\rho_\sigma(p) = p^\sigma$ is an
  homomorphism when the concatenation operation is considered.
\end{proposition}

\noindent \emph{Proof} \\
$\square$We have to check the following three properties:
\begin{enumerate}
\item $\rho_\sigma(\mathbf{1}) = \mathbf{1}$.
\item $\rho_\sigma(p_2;p_1) = \sigma;p_2;p_1;\sigma =
  \sigma;p_2;\sigma;\sigma;p_1;\sigma =
  \rho_\sigma(p_2);\rho_\sigma(p_1)$.
\item $\rho_\sigma\left( p^{-1} \right) = \sigma;p^{-1};\sigma =
  \left( \sigma^{-1};p;\sigma^{-1} \right)^{-1} = \left(
    \sigma;p;\sigma \right)^{-1} = \left[ \rho_\sigma\left( p \right)
  \right]^{-1}$. \proofend
\end{enumerate}

\section{Formal Grammar}
\label{sec:formalGrammar}

In this section we give a definition of MGG as a formal grammar and
postpone its study as a model of computation to the next
section. 

In this author's opinion, the main drawback to approach the
\textbf{P}vs\textbf{NP} problem (complexity theory in general) is the
lack of branches of mathematics available.\footnote{An interesting
  exception being that of GCT by Mulmuley and Sohoni
  (\cite{Mulmuley}).} The motivation to introduce MGG as a formal
grammar is to pose an algebraic approach to complexity theory. This
is done prior to its definition as a model of computation to make the
link more evident.

The exposition moves from the abstract to the concrete. Recall that a
formal grammar is the quad-tuple $\left( \mathcal{N}, \Sigma,
  \mathcal{P}, s_0 \right)$, being $\mathcal{N}$ a finite set of
non-terminal symbols, $\Sigma$ a finite set of terminal symbols
$\left( \mathcal{N} \cap \Sigma = \emptyset \right)$, $\mathcal{P}$ a
finite set of production rules and $s_0 \in \mathcal{N}$ the start
symbol. Production rules have the general form
\begin{equation}
  \label{eq:1}
  \left( \Sigma \cup \mathcal{N} \right)^* \mathcal{N} \left( \Sigma
    \cup \mathcal{N} \right)^* \longrightarrow \left( \Sigma \cup
    \mathcal{N} \right)^*,
\end{equation}
where ${}^*$ is the Kleene star operator.\footnote{If $V$ is a set
  then $V^*$ is the smallest superset that contains the empty string
  and is closed under the string concatenation operation.} See for
example~\cite{Hopcroft}. Equation~\eqref{eq:1} demands at least one
non-terminal symbol on the LHS. The operations on chains of symbols
are concatenation and non-terminal symbol replacement, the latter
defined through production rules. We shall dedicate the rest of the
section to these two operations.


We shall start with the set of $n \times n$ matrices over the field
$\mathbb{Z}_2[i]$, $M_n \left( \mathbb{Z}_2[i] \right)$. An element $z
= x + iy \in \mathbb{Z}_2[i]$ is such that $x, y \in \{0,1\}$. We
shall consider the following operations over $M_n \left(
  \mathbb{Z}_2[i] \right)$:
\begin{enumerate}
\item \emph{Multiplication by scalars}. Represented by $\cdot$. As
  we are in $\mathbb{Z}_2$, this operation either lets the element
  unaltered or transforms it into the zero matrix $\mathbf{0}$.
\item \emph{Matrix addition}. Represented by $+$ and defined in the
  usual way.
\item \emph{Matrix multiplication}. Represented by $\odot$ or by
  $\cdot$ if no confusion with the multiplication by scalars may rise.
  Defined as usual, with addition and multiplication\footnote{In MGG,
    the matrix multiplication uses \textbf{and} and \textbf{or}. The
    \textbf{and} operation coincides with the multiplication over
    $\mathbb{Z}_2$. It is not difficult to see that the \textbf{or}
    operation can be replaced by the \textbf{xor} as long as one of
    the matrices involved is a permutation matrix. The equality $p
    \vee q = p + pq + q$ in propositional logics can be used together
    with the fact that all elements but one in a row (column) are
    zero.} carried out over $\mathbb{Z}_2$.
\item \emph{Pointwise multiplication}. Represented by $\wedge$
  (omitted by default). Let be given $\mathcal{L}, \mathcal{R} \in M_n
  \left( \mathbb{Z}_2[i] \right)$ with $\mathcal{L} = L + iK$ and
  $\mathcal{R} = R + iQ$, then
  \begin{equation}
    \label{eq:26}
    \mathcal{L} \wedge \mathcal{R} = \mathcal{L} \mathcal{R} = (LR +
    KQ) + i(LQ + KR)
  \end{equation}
  where $XY$ is the elementwise multiplication, $XY = \left( x^j_m
    y^j_m \right)_{j, m \in \{1, \ldots, n\}}$.
\item \emph{Scalar product}. Let $\mathcal{L}_j = L_j + i K_j \in
  M_n \left( \mathbb{Z}_2[i] \right)$, then the scalar product is
  represented by $\left \langle \mathcal{L}_1, \mathcal{L}_2 \right
  \rangle$ and defined if
  \begin{equation}
    \label{eq:27}
    (L_1 + K_1)(\mathbf{1} + L_2 + K_2) = 0.
  \end{equation}
  This condition guarantees that $\mathcal{L}_1, \mathcal{L}_2 \in M_n
  \left( \mathbb{Z}_2[i] \right) \implies \left \langle \mathcal{L}_1,
    \mathcal{L}_2 \right \rangle \in M_n \left( \mathbb{Z}_2[i]
  \right)$. See the proof of Prop.~\ref{prop:MisClosed} on
  p.\pageref{prop:MisClosed}.
\end{enumerate}

As we shall be almost exclusively interested in those elements that
have disjoint real and imaginary parts, let's introduce
\begin{equation}
  \label{eq:2}
   \widetilde{\mathbb{M}}^n = \left\{ \mathcal{L} = L + iK \in M_n
     \left( \mathbb{Z}_2[i] \right) \; \left| \; L K = 0
     \right. \right\}.
\end{equation}

Elements of $\widetilde{\mathbb{M}}^n$ will be known as \emph{Boolean
  complex matrices}. Let $\mathcal{T}$ be some finite set and for
$\tilde{g} \in \widetilde{\mathbb{M}}^n$ consider the mapping
$\lambda_{\tilde{g}}: \tilde{g} \rightarrow \mathcal{T}$ defined only
for elements in the diagonal of $L$ and $K$ and such that $\lambda(L)
= \lambda(K)$. The meaning of $\lambda$ will be clarified in
Sec.~\ref{sec:modelOfComputation}. Its subscript will be usually
omitted.

Except for some restrictions on the operations -- their concrete
definition taking $\lambda$ into account are given in
eqs.~(\ref{eq:8}),~(\ref{eq:7}) and~(\ref{eq:5}) below -- the (vector)
space on which we shall focus our attention is $\left( \mathbb{M}^n,
  +, \cdot\,, \odot \right)$, where
\begin{equation}
  \label{eq:9}
  \mathbb{M}^n = \left\{ g = \left( \tilde{g}, \lambda \right) \;
  \left| \; \tilde{g} \in \widetilde{\mathbb{M}}^n \textrm{ and }
    \lambda \textrm{ as above} \right. \right\}.
\end{equation}

For MGG, the quad-tuple $\mathfrak{G} = \left( \mathcal{N}, \Sigma,
  \mathcal{P}, s_0 \right)$ that defines the formal grammar has the
following elements:
\begin{itemize}
\item $\mathcal{N} = \left\{ \mathcal{L}_i \; \vert \; \exists p_i:
    \mathcal{L}_i \rightarrow \mathcal{R}_i, \; p_i \in \mathcal{P}
  \right\}$, the finite set of elements in $\mathbb{M}^n$ that appears
  in the LHS of some production rule.
\item $\Sigma = \left\{ \mathcal{L} \in \mathbb{M}^n \; \vert \;
    \mathcal{L} \not \in \mathcal{N} \right\}$, the finite set of
  elements in $\mathbb{M}^n$ not belonging to $\mathcal{N}$.
\item $\mathcal{P} = \left\{ p: \mathcal{L} \rightarrow \mathcal{R} \;
    \vert \; \mathcal{L} \in \mathcal{N}, \; \mathcal{R} \in \Sigma
    \cup \mathcal{N} \right\}$, the finite set of production rules as
  defined in eq.~\eqref{eq:4} below.
\item $S_0 \in \mathbb{M}^n$, the start symbol.
\end{itemize}

MGG as a formal grammar (and as a model of computation) will be
limited to ``affine'' mappings, which correspond to non-terminal
symbol replacement (concatenation is addressed by the end of the
section). A grammar rule $p$ has the general form
\begin{equation}
  \label{eq:4}
  \mathcal{S}_1 = p(\mathcal{S}_0) = \left \langle S_0, \sigma . \,
    \omega \right \rangle = \left \langle \mathcal{S}_0, \sigma \cdot
    \omega \cdot \sigma^t \right \rangle,
\end{equation}
where $\mathcal{S}_0, \mathcal{S}_1 \in \mathbb{M}^n$, $\sigma$ is a
permutation matrix (a single $1$ per row and column; see
Sec.~\ref{sec:relabeling}) and $\omega \in \mathbb{M}^n$ is such that
there exists $\alpha \in \mathbb{M}^n$ with zero imaginary part that
fulfills
\begin{equation}
  \label{eq:6}
  \omega = \alpha + i (\alpha + \mathbf{1}),
\end{equation}
being $\mathbf{1} \in \mathbb{M}^n$ the $n \times n$ matrix filled up
with ones (by eq. \eqref{eq:2} the imaginary part must be zero). These
elements have been introduced in~\cite{MGG_MCL}, so-called
\emph{swaps}.

The way operations in eqs.~\eqref{eq:4}~and~\eqref{eq:6} are carried
out need to be clarified regarding the $\lambda$ mappings, which are
part of the elements of $\mathbb{M}^n$ according to~\eqref{eq:9}:
\begin{enumerate}
\item Multiplication by scalars is defined by
  \begin{equation}
    \label{eq:8}
    \alpha \mathcal{L} = \left( \alpha L + i \alpha K,
      \lambda(\mathcal{L}) \right), \qquad \alpha = 0, 1.
  \end{equation}
\item The addition $\mathcal{L}_1 + \mathcal{L}_2$ should respect the
  mapping $\lambda$ so it is allowed just in case
  $\lambda_{\mathcal{L}_1} (\mathcal{L}_1) =
  \lambda_{\mathcal{L}_2}(\mathcal{L}_2)$:
  \begin{align}
    \label{eq:7}
    \mathcal{L}_1 + \mathcal{L}_2 &= \left( (L_1 + L_2) + i (K_1 +
      K_2),
      \lambda(\mathcal{L}_1) \right) = \\
    &= \left( (L_1 + L_2) + i (K_1 + K_2), \lambda(\mathcal{L}_2)
    \right). \nonumber
  \end{align}
\item Recall from Sec.~\ref{sec:relabeling} that permutation matrices
  define a permutation on the indices of the elements of
  $\mathbb{M}^n$, $\sigma(1 \; 2 \; \ldots \; n) = (\sigma(1) \;
  \sigma(2) \; \ldots \; \sigma(n))$. For $\mathcal{L} \in
  \mathbb{M}^n$ the matrix multiplication is given by
  \begin{align}
    \label{eq:5}
    \sigma . \, \mathcal{L} = \sigma \cdot \mathcal{L} \cdot \sigma^t
    &= \sigma \cdot (L + iK, \lambda) \cdot \sigma^t = \nonumber \\
    &= \left( \sigma \cdot (L + i K ) \cdot \sigma^t, \sigma \left(
        \lambda \left( \mathcal{L}_{jj} \right)
      \right) \sigma^t \right) = \nonumber \\
    &= \left( \sigma \cdot L \cdot \sigma^t + i \sigma \cdot K \cdot
      \sigma^t, \lambda \left( \mathcal{L}_{\sigma(j)\sigma(j)}
      \right) \right).
  \end{align}
\end{enumerate}

The following proposition now follows easily:

\begin{proposition}
  \label{prop:MisClosed}
  The formal grammar $\mathfrak{G}$ is closed under the operations
  given by the production rules as defined in eq. \eqref{eq:4}.
\end{proposition}

\noindent \emph{Proof} \\
$\square$The proof can almost be derived from the fact that
$\mathbb{M}^n$ is closed under the operations $\cdot$, $+$, $\odot$
defined in eqs. \eqref{eq:8}, \eqref{eq:7} and \eqref{eq:5}:
\begin{enumerate}
\item $\alpha = 0,1, \; \mathcal{L} \in \mathbb{M}^n \implies \alpha
  \cdot \mathcal{L} \in \mathbb{M}^n$ holds because $\mathbf{0} \in
  \mathbb{M}^n$.
\item $\mathcal{L} \in \mathbb{M}^n, \sigma \textrm{ permutation}
  \implies \sigma \cdot \mathcal{L} \cdot \sigma^t \in \mathbb{M}^n$
  because it is just a relabeling.
\item $\mathcal{L}_1, \mathcal{L}_2 \in \mathbb{M}^n \implies L_1 +
  L_2 \in \mathbb{M}^n$ because addition respects labeling.
\end{enumerate}
A little of extra work is needed for the scalar product:
\begin{align*}
  \left \langle \mathcal{L}_1, \mathcal{L}_2 \right \rangle &= \left
    \langle L_1 + i K_1, L_2 + i K_2 \right \rangle = \left( L_1 + i
    K_1 \right) \left( \mathbf{1} + K_2 + i (\mathbf{1} + L_2) \right)
  = \\
  &= \left( L_1 + L_1 K_2 + K_º + K_1 L_2 \right) + i \left( K_1 + K_1
  K_2 + L_1 + L_1 L_2 \right) = \\
&= \alpha + i \beta.
\end{align*}
so $\left \langle \mathcal{L}_1, \mathcal{L}_2 \right \rangle \in
\mathbb{M}^n$ if $\alpha \beta = 0$. After some manipulations, we
obtain that
\begin{equation}
  \label{eq:29}
  \alpha \beta = \left( L_1 + K_1 \right) \left( \mathbf{1} + L_2 +
    K_2 \right).
\end{equation}
To see that eq. \eqref{eq:6} guarantees $\alpha \beta = 0$ simply
substitute it into eq. \eqref{eq:29}. \proofend

Concatenation as an operation is simpler than non-terminal symbol
replacement. In essence it consists in passing from $\mathbb{M}^{m}$
to $\mathbb{M}^{n}$ with $m \leq n$. Assuming that $\mathcal{T}$ is
not enlarged,\footnote{If $\mathcal{T}$ is modified then we would be
  redefining the grammar rather than concatenating symbols.} all we
have to do is to add $n - m$ rows and columns to the matrices that
define the production rules (filling them with zeros) and consider
these new rows and columns in the permutation matrices.

The production rules in $\mathfrak{G}$ have the general form
$\mathcal{N} \rightarrow \Sigma \cup \mathcal{N}$, which is a
particular case of eq. (\ref{eq:1}). Taking concatenation into
account, the general form that appears in eq. (\ref{eq:1}) is
recovered.

\section{Model of Computation}
\label{sec:modelOfComputation}

In this section we shall start by informally providing some semantics
to the elements and operations of the formal grammar $\mathfrak{G}$ of
Sec.~\ref{sec:formalGrammar}: elements of $\mathbb{M}^n$ (permutation
matrices in particular), addition, matrix multiplication, the set
$\mathcal{T}$, the mapping $\lambda$ and the element $\omega$ as
introduced in eq. (\ref{eq:6}). After this, the nodeless MGG model of
computation is defined. The section is closed with a short summary on
some MGG submodels and supermodels of computation.

Let $\mathcal{L} = L + iK \in \mathbb{M}^n$. The Boolean matrix $L$
can be seen as the adjacency matrix of some simple digraph $g$. This
matrix is completed with $K$, which we will interpret as those edges
that can not belong to $g$.\footnote{The importance of $K$ stems from
  the fact that we will be studying graph dynamics. We need to specify
  that one edge is not present in a given graph for example because it
  is going to be added by some production.}

A meaning for $\mathcal{T}$ could be that of (node) labels. Closely
related is the mapping $\lambda$ that assigns a label to each
node. This is why it is just defined for the elements in the main
diagonal. It is natural to impose $\lambda(L) = \lambda(K)$. The
operation defined by eq. (\ref{eq:5}) can be understood as a node
relabeling. Refer to Sec.~\ref{sec:relabeling}.

The definition of $\omega$ given in eq. (\ref{eq:6}) simply states
that we have to do inverse operations on $L$ and $K$ (refer to Sec.
\ref{sec:MGGbasics}). The permutation matrix $\sigma$ allows $\omega$
to act on rearrangements of elements equally labeled.

Notice that no restrictions can be set on the applicability of
production rules according to eq. (\ref{eq:4}), i.e. swaps can always
be applied. We shall introduce a new means to define productions (as
introduced in MGG; see Sec.~\ref{sec:MGGbasics}) using the Boolean
operations \textbf{and} (pointwise multiplication) and \textbf{or}
(close to matrix addition) and the erasing and addition matrices. This
will be our model of computation, to be known as \emph{nodeless MGG}.
Productions can be understood as swaps with restrictions: productions
are one of the possible ways to set constraints on the applicability
of swaps.\footnote{We shall see in Sec.~\ref{sec:BA2VS} how swaps and
  productions as introduced here are related.}

The model of computation associated to MGG is the 5-tuple $\left(
  \mathcal{G}, \mathcal{T}, \mathcal{P}, \tau, H_0 \right)$ where
\begin{itemize}
\item $\mathcal{G} = \left\{ \mathcal{L} = (L, K, \lambda) \;
    \left\vert \right. \; \lambda: V(\mathcal{L}) \rightarrow
    \mathcal{T} \right\}$, being $V(\mathcal{L})$ the set of
  vertices. The formal notation $\mathcal{L} = L \vee iK$ will be
  used.
\item $\mathcal{T}$ is the finite set of labels.
\item $\mathcal{P}$ is the finite set of compatible productions.
\item $\tau$ is the transition function that modifies the
  state:\footnote{\emph{Grammar state} will be for us the next
    production to apply and where it is to be applied (match of the
    production in the host graph). In some sense, the state of the
    productions. With \emph{system state} we shall refer to the actual
    host graph (the state of the object under study). The term
    \emph{state} alone will mean the grammar state plus the system
    state.}
  \begin{align*}
    \tau: \mathcal{G} \times \mathcal{P} &\longrightarrow \mathcal{G}
    \\
    (H, p) & \longmapsto \langle H, p \rangle = p(H) = H'
  \end{align*}
\item $H_0$ is the initial system state (see
  Sec.~\ref{sec:MGGbasics}).
\end{itemize}

The model is deterministic if $\tau$ is a single-valued function and
non-deterministic if it is a multivalued function. The default halting
condition is ``no production can be applied''.

The two basic operations are again non-terminal symbol replacement and
concatenation, on which we touch in the following paragraphs.

Concatenation in MGG is defined for productions, which is just the
sequential application of two or more productions. It will be
represented by a semicolon and should be read from right to left (like
standard composition of functions). Refer to Sec.~\ref{sec:MGGbasics}
for more details.

Graph rewriting substitutes the occurrence of some graph (known as
pattern graph or production LHS, $L$) in the host graph $(H)$ by the
corresponding replacement graph (also known as production RHS, $R$).
The specification of the operation is done through a so-called
\emph{grammar rule}, \emph{production rule} or just \emph{production},
and is represented as a graph morphism $p: L \rightarrow R$. The
operation itself is known as a \emph{derivation}. Refer again to Sec.
\ref{sec:MGGbasics}. Graph rewriting plays in MGG the role of
non-terminal symbol replacement in formal grammars.

We can associate the element $p = e \vee i r$ to any production $p$,
being $e$ and $r$ their erasing and addition matrices, respectively,
which have been introduced in Sec.~\ref{sec:MGGbasics}. Notice that as
$e$ and $r$ are disjoint, the \textbf{or} and the matrix addition
coincide, so we can think of $p$ as an element of $\mathbb{M}^n$.

From the element $p = e \vee i r$ we can define a swap $\omega = P(p)
= P(e \vee i r) = \overline{e} \, \overline{r} \vee i(e \vee
r)$. Again, as $\overline{e} \, \overline{r}$ and $e \vee r$ are
disjoint we can write
\begin{equation}
  \label{eq:33}
  \omega = (e + \mathbf{1}) (r + \mathbf{1}) + i (e + r) = (e + r +
  \mathbf{1}) + i (e + r).
\end{equation}
Notice that $\omega$ satisfies eq. \eqref{eq:6}, hence the
name. Moreover, as was proved in \cite{MGG_Combinatorics}, $R = \left
  \langle L, P(p) \right \rangle$. Taking relabeling into account we
can write an affine production as
\begin{align}
  R \vee i Q = \mathcal{R} = p(\mathcal{L}) &= \sigma \cdot \left
    \langle \sigma \cdot ( L \vee i K ) \cdot \sigma^t, \overline{e}
    \, \overline{r} \vee i \left( e \vee r \right) \right \rangle
  \cdot \sigma^t = \nonumber \\
  &= \left \langle L \vee i K, \sigma \cdot \left[ \overline{e}
      \, \overline{r} \vee i \left( e \vee r \right) \right] \cdot
    \sigma^t \right \rangle,
\label{eq:28}
\end{align}
which is just eq.~\eqref{eq:31} including the nihil parts $K$ and
$Q$.

The application of the production $p$ to a host graph $G$ to derive
the system state $H$ is given by:
\begin{equation}
  \label{eq:44}
  H = p(G) = \left \langle G, \sigma . \, P(p) \right \rangle.
\end{equation}
The matching -- see Chap.~6 in~\cite{MGG_Book} -- is performed by
the relabeling $\sigma$. Notice the non-determinism of this step and
its \textbf{NP}-completeness. Compare with eq.~\eqref{eq:4}. 


\begin{figure}[htbp]
  \centering
  \includegraphics[scale = 0.6]{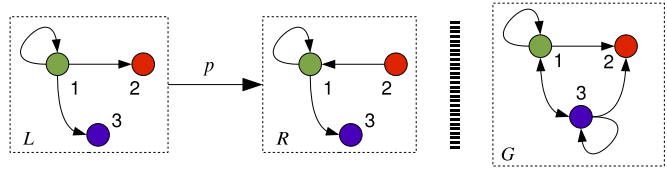}
  \caption{Sample Permutation plus Host Graph}
  \label{fig:relabelingExample}
\end{figure}

\noindent \textbf{Example}.$\square$Let's consider the
production $p$ and the host graph $G$ depicted in
Fig.~\ref{fig:relabelingExample}. The nihil terms $K$ and $Q$ have
been omitted to ease exposition. All nodes are assumed to be of the
same type so neither the numbers nor the colors in the figure should
be understood as labels. Numbers are used for referencing purposes
only.

Production $p$ deletes edge $(1,2)$ and adds a new edge $(2,1)$. It
also keeps edges $(1,1)$ and $(1,3)$, demanding their existence in the
host graph. There are initially four possible identifications of the
LHS $L$ in $G$, which correspond to the following mappings: $(1, 2, 3)
\longmapsto (1,2,3), (1,3,2), (3,2,1)$ and $(3,1,2)$. The permutation
matrices are
\begin{displaymath}
  \sigma_1 = \left[ \begin{array}{ccc}
      \vspace{-4pt}
      1 & 0 & 0 \\
      \vspace{-4pt}
      0 & 1 & 0 \\
      \vspace{-4pt}
      0 & 0 & 1 \\
      \vspace{-10pt}
    \end{array} \right] \quad \sigma_2 = \left[ \begin{array}{ccc}
      \vspace{-4pt}
      1 & 0 & 0 \\
      \vspace{-4pt}
      0 & 0 & 1 \\
      \vspace{-4pt}
      0 & 1 & 0 \\
      \vspace{-10pt}
    \end{array} \right] \quad \sigma_3 = \left[ \begin{array}{ccc}
      \vspace{-4pt}
      0 & 0 & 1 \\
      \vspace{-4pt}
      0 & 1 & 0 \\
      \vspace{-4pt}
      1 & 0 & 0 \\
      \vspace{-10pt}
    \end{array} \right] \quad \sigma_4 = \left[ \begin{array}{ccc}
      \vspace{-4pt}
      0 & 0 & 1 \\
      \vspace{-4pt}
      1 & 0 & 0 \\
      \vspace{-4pt}
      0 & 1 & 0 \\
      \vspace{-10pt}
    \end{array} \right]
\end{displaymath}
with associated productions
\begin{equation*}
  q_i(L) = \langle L, \sigma_i.\,p  \rangle = \langle L, \sigma_i
  \cdot p \cdot \sigma_i^t \rangle, \quad i \in \{1,2,3,4\}.
\end{equation*}
In principle, all four are valid productions but not all can be
applied to $G$. Production $p_2$ would try to add edge $(3,1)$ which
already exists in $G$. The same problem appears with $p_4$ but this
time with edge $(1,3)$.

Apart from illustrating how relabeling works on productions, this
example tries to show how it will be of help in characterizing
non-determinism in Sec.~\ref{sec:nonDeterminism}. \proofend

The main difference between what is exposed in this section and
Sec.~\ref{sec:MGGbasics} is that nodeless MGG does not act on nodes
but just on edges. This has the advantage of avoiding dangling edges
and all compatibility issues derived.



The section ends brushing over submodels (by setting further
constraints on MGG) and supermodels of computation (by relaxing the
\emph{axioms} of nodeless MGG). An initial proposal
is:
\begin{enumerate}
\item On the matching:
  \begin{itemize}
  \item For submodels, a \emph{subisomorphism}\footnote{A
      \emph{subisomorphism} can be defined as an isomorphism between
      the LHS of a production and the part of the host graph in which
      it is identified.}  instead of an injective morphisms -- as in
    Def.  \ref{def:directDerivationDef} -- can be demanded.
  \item As a supermodel, the injectivity of the morphism may be
    relaxed.
  \end{itemize}
\item On the productions:
  \begin{itemize}
  \item For submodels, relabeling can be forbidden. Other constraints
    on the operations can be set by using application conditions
    (see~\cite{MGG_Book}, Chaps.~8~and~9).
  \item Nodeless MGG can be extended by allowing graph constraints and
    application conditions (see~\cite{MGG_Book}). Also, new operations
    on graphs can be allowed such as negation of graphs or we can
    permit general adjacency matrices in the multiplication instead of
    just permutational matrices (see
    Prop.~\ref{prop:permsPerSwapAndProd} below).
  \end{itemize}
\item On the underlying space:
  \begin{itemize}
  \item For submodels, instead of considering simple digraphs we may
    consider undirected graphs or even subsets such as the group of
    permutations (digraphs with single incoming and outgoing edges).
  \item Other structures more general than simple digraphs can be
    allowed. Examples are multidigraphs and hypergraphs.
  \end{itemize}
\end{enumerate}

Limiting matrix multiplication to permutation matrices is a big
restriction, at least concerning the amount of matrices left out. The
following proposition shows the orders as functions on the number of
nodes:
\begin{proposition}
  \label{prop:permsPerSwapAndProd}
  Assuming that all nodes are of the same type, the orders of the
  number of permutation matrices per swap $P_\omega(n)$ and per
  production $P_p(n)$ are:
  \begin{align}
    \label{eq:68}
    P_\omega(n) = \frac{\#(\mathtt{perms})}{\#(\mathtt{swaps})} =
    \frac{n!}{2^{n^2}} &\sim \frac{\sqrt{\pi n} \left( \frac{n}{e}
      \right)^n}{2^{n^2 - 1/2}} \in e^{- O \left( n^2 \right)}, \\
    \label{eq:25}
    P_p(n) = \frac{\#(\mathtt{perms})}{\#(\mathtt{prods})} =
    \frac{n!}{3^{n^2}} &\sim \frac{\sqrt{2 \pi n} \left( \frac{n}{e}
      \right)^n}{3^{n^2}} \in e^{-O \left( n^2 \right)},
  \end{align}
  where $n$ is the number of nodes.
\end{proposition}

\noindent \emph{Proof} \\
$\square$Recall that $\log_b(x) = \frac{\log_c(x)}{\log_c(b)}$. The
Stirling formula for the factorial (the number of permutations) has
been used in eqs. \eqref{eq:68} and \eqref{eq:25}. For the
denominator, notice that a swap is fixed by its real or imaginary
parts, so there are as many as adjacency matrices for graphs with $n$
nodes: $2^{n^2}$. To prove eq.~\eqref{eq:68} just take the logarithms
to derive
\begin{equation}
  \label{eq:32}
  \frac{\log(2) + \log{\pi}}{2} + \frac{\log(n)}{2} - n + n \log(n) -
  \log(2) n^{2}.
\end{equation}

\begin{figure}[htbp]
  \centering
  \includegraphics[scale = 0.75]{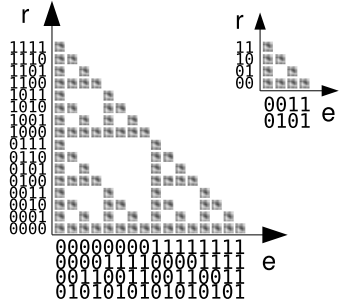}
  \caption{Weel-Defined Productions}
  \label{fig:imageP}
\end{figure}

All we have to do to check eq. \eqref{eq:25} is to count the number of
different productions that can be defined for graphs with $n$
nodes. One possibility is to establish a link between productions and
the standard Sierpinski gasket.\footnote{For more on the relationship
  between the Sierpinski gasket and swaps and productions, please
  refer to~\cite{MGG_MCL}.} To this end, write the adjacency matrix of
the graph $g$ of $n$ nodes as a single column (the second sitting
right after the first, the third after the second and so on). Recall
that a production $p$ can be written as $p = e \vee i r$, with $er =
0$. Hence, productions are the zeros of the \textbf{and} function. See
Fig.~\ref{fig:imageP} which represents well defined productions of the
form $p = e \vee ir$ for matrices with two nodes.

The Lucas correspondence theorem (see \cite{fine}~and
also~\cite{Weisstein}) proves that the set of zeros is the Sierpinski
gasket as it can be used to compute the binomial coefficient
$\binom{L}{K} \mod 2$ with bitwise operations: $\overline{L} \wedge
K$. This tells us that the parity of the function $\binom{L}{K}$ (this
is what the function $\!\!\!\!\mod 2$ does) is the same as that of
$\overline{L} \wedge K$. In our case $L$ is the abscissa. The negation
of $L$ just reverts the order (it is a symmetry) and does not change
the shape of the figure. 

Counting the number of elements is not difficult. Notice that in
Fig.~\ref{fig:imageP}, the big Sierpinsky gasket is made of a number
of copies of the smaller one. The small Sierpinsky gasket has $9$
elements and it appears $3^{n^2-2}$ times so the total amount of well
defined productions is $3^{n^2}$. The main difference with respect to
eq.~(\ref{eq:32}) is the coefficient of $-n^2$ which would be
$\log(3)$ this time. \proofend

\section{Determinism and Non-Determinism in MGG}
\label{sec:nonDeterminism}

There are two types (sources) of non-determinism associated to the
grammar state: selecting the next production to apply (let's call it
\emph{election non-determinism}) and finding the place in the host
graph where the production will be applied (that we shall name
\emph{allocation non-determinism}). The transition function is not
uniquely determined either because more than one production can be
applied or because the oracle returns more than one place where the
production can be applied.

\begin{figure}[htbp]
  \centering
  \includegraphics[scale = 0.46]{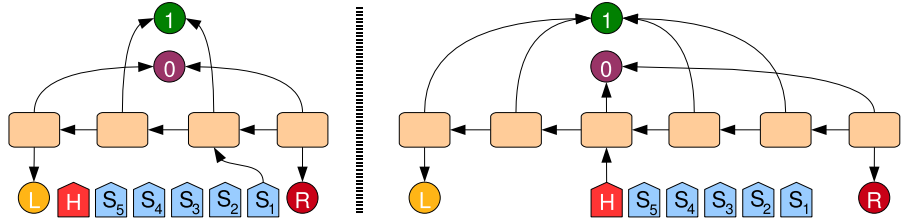}
  \caption{Initial and Final States for the Copy Subroutine}
  \label{fig:TMinitialState}
\end{figure}

One easy way to mitigate or to even remove non-determinism is to use
\emph{control nodes} that indicate what production to apply or where
it should be applied. Examples of these control nodes are the $S_i$
that appear in the productions of Sec.~\ref{sec:MGGandTM},
Figs.~\ref{fig:TMprods1},~\ref{fig:TMprods2}~and~\ref{fig:TMinitialState}. It
is also possible that, for certain purposes, which production is
applied or where it is applied becomes irrelevant.\footnote{This is
  related to \emph{confluence}, not addressed in the present
  contribution. See~\cite{termRewriting_Book}.} An example of this
behavior is the simulation of BCs given in Sec.~\ref{sec:MGGandBC}.

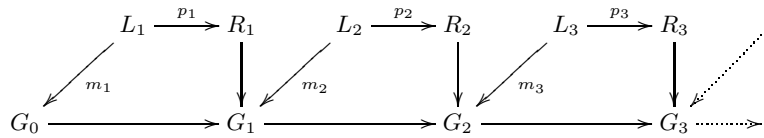
\begin{figure}[htbp]
  \centering \makebox{ \xymatrix{
      & L_1 \ar[dl]^{m_1} \ar@[blue][r]^{p_1} & R_1 \ar[d] & L_2
      \ar[dl]^{m_2} \ar@[blue][r]^{p_2} & R_2 \ar[d]& L_3
      \ar[dl]^{m_3} \ar@[blue][r]^{p_3} & R_3 \ar[d] & \ar@{.>}[dl] \\
      G_0 \ar@[red][rr] && G_1 \ar@[red][rr] && G_2 \ar@[red][rr] &&
      G_3 \ar@{.>}@[red][r] &
    } }
  \caption{Productions Application (Derivation: Evolution of a
    System)}
  \label{fig:sequence}
\end{figure}

If a sequence of productions $s = p_n; \ldots; p_1$ (see
Fig.~\ref{fig:sequence}) is considered instead of a single production,
allocation non-determinism can be graded. There is a first
\emph{partial} level where only productions are taken into account
(ignoring the initial and intermediate system states) and a second
complete level if the host graphs are considered. Let's call
them \emph{horizontal non-determinism} and \emph{vertical
  non-determinism}, respectively.

\noindent \textbf{Remark}.$\square$The names stem from the way the
identification of elements is performed according to the
representation in Fig.~\ref{fig:sequence}. If the system states are
not taken into account, the identification of nodes is horizontal (in
productions $p_i$). If the system states are considered, the
identifications are given by the matching mappings $m_i$. \proofend

Nodes with the same label are potentially interchangeable so
non-determinism in a single graph is equivalent to the Cartesian
product of the corresponding permutation groups (one per type). For a
given simple digraph $L$, let's denote this group by $S(L)$. Their
elements are represented by permutation matrices as explained in
Sec.~\ref{sec:relabeling}.

An affine production $p:L \rightarrow R$ identifies nodes between $L$
and $R$ in a unique way. The traslational part of the production that
acts on edges does not affect the permutation group of the RHS. The
rotational part $\sigma$ does not affect the permutation group of the
RHS either:
\begin{equation*}
  \mathcal{S}(R) = \{ \sigma \cdot \sigma_i \, \vert \, \sigma_i \in
  \mathcal{S}(L)\} = S(L).
\end{equation*}
Hence, productions in nodeless MGG\footnote{Productions would change
  the permutation groups if they were allowed to add or remove nodes.}
do not modify the permutation group associated to graphs.

Horizontal non-determinism in sequences can be handled by letting the
permutational part of productions vary in the corresponding
permutation group. Notice however that not all variations are possible
if applicability is to be kept (see the example of
Fig.~\ref{fig:relabelingExample}).




\begin{figure}[htbp]
  \centering
  \includegraphics[scale = 0.5]{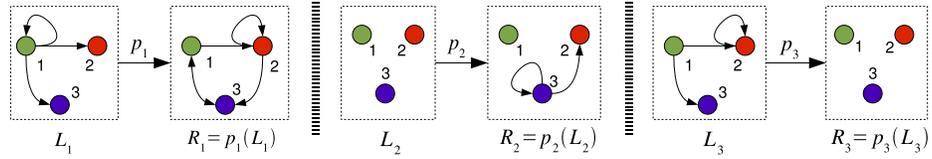}
  \caption{Sample Productions}
  \label{fig:SeqNonDet}
\end{figure}

\noindent \textbf{Example}.$\square$Let's consider the productions
$p_1$, $p_2$ and $p_3$ depicted in Fig.~\ref{fig:SeqNonDet}, being
$1$, $2$ and $3$ nodes of the same type. The group of permutations of
three elements is $S_3 = \{id, (1 \; 2), (1 \; 3), (2 \; 3), (1 \; 2
\; 3), (3 \; 2 \; 1)\}$. Their associated affine productions are
\begin{equation*}
  q^\sigma_i(L_i) = \left \langle L_i, \sigma. \, p_i \right \rangle =
  \left \langle L_i, \sigma \cdot p_i \cdot \sigma^t \right \rangle,
  \quad i \in \{1,2,3\},
\end{equation*}
with $\sigma \in S_3$. The sequence $s = p_3;p_2;p_1$ has associated
affine sequence
\begin{equation*}
  s' = q_3; q_2; q_1 =  \langle  \langle \langle L_1,
  \sigma^1\!\!.\,p_1 \rangle, \sigma^2\!\!.\,p_2 \rangle,
  \sigma^3\!\!.\,p_3 \rangle.
\end{equation*}

\begin{table*}[htbp]
  \centering
  \begin{tabular}{|c|c|c|c|c|c|c|}
    \hline
    \phantom{$I^{I^I}$}$\sigma^2 \backslash \sigma^1$\phantom{$I^{I^I}$} &
    $\;\;\;id\;\;\;$ & $\,\;(1 \; 2)\;\,$ & $\,\;(1 \; 3)\;\,$ &
    $\,\;(2 \; 3)\;\,$ & $\,(1 \; 2 \; 3)\,$ & $\,(3 \; 2 \; 1)\,$ \\
    \hline
    $id$ & 1 & 0 & 0 & 0 & 0 & 0 \\
    \hline
    $(1 \; 2)$ & 0 & 1 & 0 & 0 & 0 & 0 \\
    \hline
    $(1 \; 3)$ & 0 & 0 & 1 & 0 & 0 & 0 \\
    \hline
    $(2 \; 3)$ & 0 & 0 & 0 & 1 & 0 & 0 \\
    \hline
    $\,(1 \; 2 \; 3)\,$ & 0 & 0 & 0 & 0 & 1 & 0 \\
    \hline
    $\,(3 \; 2 \; 1)\,$ & 0 & 0 & 0 & 0 & 0 & 1 \\
    \hline
  \end{tabular}
  \caption{Possible Relabelings for Sequence $s = p_2;p_1$}
  \label{tab:relsForTwoProductions}
\end{table*}

Table~\ref{tab:relsForTwoProductions} includes all the permutations
for the subsequence $p_2;p_1$. A $1$ means that applicability is kept
and a $0$ means that applicability is not kept. Only the diagonal
keeps applicability. The problem in all cases is that some already
existing edge would be added by $q_2$.

\begin{table*}[htbp]
  \centering
  \begin{tabular}{|c|c|c|c|c|c|c|}
    \hline
    \phantom{$I^{I^I}$}\!\! $\sigma^3 \backslash \sigma^2 \sigma^1$
    \phantom{$I^{I}$} & $[id,\!id]$ & $[(1 \, 2),\!(1 \, 2)]$ & $[(1 
    \, 3),\!(1 \, 3)]$ & $[(2 \, 3),\!(2 \, 3)]$ & $[(1 \, 2 \,
    3),\!(1 \, 2 \, 3)]$ & $[(3 \, 2 \, 1),\!(3 \, 2 \, 1)]$
    \\
    \hline
    $id$ & 1 & 0 & 1 & 1 & 0 & 0 \\
    \hline
    $(1 \; 2)$ & 0 & 1 & 0 & 0 & 1 & 1 \\
    \hline
    $(1 \; 3)$ & 1 & 0 & 1 & 0 & 0 & 1 \\
    \hline
    $(2 \; 3)$ & 1 & 0 & 0 & 1 & 1 & 0 \\
    \hline
    $\,(1 \; 2 \; 3)\,$ & 0 & 1 & 0 & 1 & 1 & 0 \\
    \hline
    $\,(3 \; 2 \; 1)\,$ & 0 & 1 & 1 & 0 & 0 & 1 \\
    \hline
  \end{tabular}
  \caption{Possible Relabelings for Sequence $s = p_3;p_2;p_1$}
  \label{tab:relsForThreeProductions}
\end{table*}

Table~\ref{tab:relsForThreeProductions} summarizes all permutations
for the sequence $p_3;p_2;p_1$. Just the permutations that keep
applicability for $p_2;p_1$ has been considered (indexed by
columns). Again, a $1$ means that applicability is kept and a $0$
means that applicability is not kept. The problem in all cases is that
some non-existent edge would be deleted by $q_3$.

\begin{figure}[htbp]
  \centering
  \includegraphics[scale = 0.52]{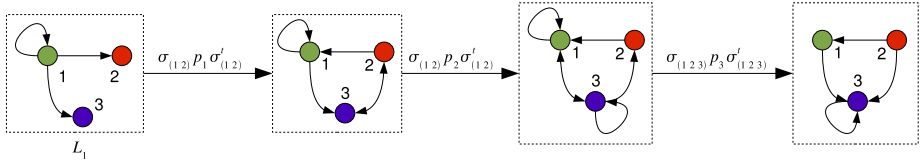}
  \caption{Image of the Sequence of Three Productions}
  \label{fig:seqThreeProds}
\end{figure}

In this example, as there are many possible relabelings, we have that
the sequence is horizontally non-deterministic. Figure
\ref{fig:seqThreeProds} represents the effect of applying the sequence
$s = q^{(1 \; 2 \; 3)}_3; q^{(1 \; 2)}_2; q^{(1 \; 2)}_1$ to the
initial host graph $L_1$. \proofend

\begin{proposition}[Horizontal Determinism Characterization]
  \label{prop:nonDet}
  Let $s_n = p_n; \ldots ;p_1$ be a compatible sequence of productions
  and define the operator $\widetilde{C}(s_n) = \widetilde{C}^+(s_n)
  \vee \widetilde{C}^-(s_n)$ by
  \begin{align}
    \label{eq:30}
    \widetilde{C}^+(s_n) &= \bigvee_{j=1}^n \left( R^\sigma_j
      \underset{j+1}{\overset{n}{\nabla'}} \left(
        \overline{e}^\sigma_x \, r^\sigma_y \right) \vee L^\sigma_j
        \underset{1}{\overset{j-1}{\bigtriangleup}} \left( e^\sigma_y
            \, \overline{r}^\sigma_x \right) \right) \\
    \label{eq:34}
    \widetilde{C}^-(s_n) &= \bigvee_{i=1}^n \left( Q^\sigma_i
      \underset{j+1}{\overset{n}{\nabla'}} \left(
        \overline{r}^\sigma_x \, e^\sigma_y \right) \vee K^\sigma_i
        \underset{1}{\overset{j-1}{\bigtriangleup'}} \left(
            \overline{e}^\sigma_x \, r^\sigma_y \right) \right),
  \end{align}
  being $X_i^\sigma = \sigma_i.\,X_i = \sigma_i \cdot X_i \cdot
  \sigma^t_i$, $X_i = e_i, r_i, L_i, R_i, K_i, Q_i$, and
  \begin{equation*}
    \underset{t_0}{\overset{t_1}{\bigtriangleup'}} \left( F(x,y)
    \right) = \bigvee_{y=t_0}^{t_1} \left( \bigwedge_{x=y}^{t_1}
      \left( F(x,y) \right) \right), \;
    \underset{t_0}{\overset{t_1}{\nabla'}} \left( F(x,y)
  \right) = \bigvee_{y=t_0}^{t_1} \left( \bigwedge_{x=t_0}^{y} \left(
      F(x,y) \right) \right).
  \end{equation*}
  There are three possibilities:
  \begin{enumerate}
  \item The sequence is not applicable if for some $i$ there exists no
    permutation such that $\widetilde{C}(s_n) = \mathbf{0}$.
  \item The sequence is horizontally deterministic if for every $i$
    there exists a single permutation such that $\widetilde{C}(s_n) =
    \mathbf{0}$.
  \item The sequence is horizontally non-deterministic if for some $i$
    there exists more than one possible permutation $\sigma$ such that
    $\widetilde{C}(s_n) = \mathbf{0}$.
  \end{enumerate}
\end{proposition}

\noindent \emph{Proof} (sketch)\\
$\square$Theorem.~6.4.4 in~\cite{MGG_Book} characterizes applicability
as equivalent to compatibility plus coherence. Applicability in a
single place is determinism and applicability in several places is
non-determinism. Compatibility is always fulfilled by nodeless MGG and
nevertheless a hypothesis of the proposition, so it just remains to
study coherence.

Equations~\eqref{eq:30}~and~(\ref{eq:34}) for productions instead of
affine productions characterizes coherence as proved in Th.~2, Sec.~5
in~\cite{MGG_Combinatorics}, or in Theorems~4.3.5 and~4.4.3
in~\cite{MGG_Book}. To finish the proof, simply add permutations to
the productions (to tackle relabeling) transforming them into affine
productions. \proofend

We shall end this section by pointing out that vertical
non-determinism can be addressed as horizontal non-determinism. The
only difference is that matrices (and thus permutation groups) will be
larger.

\section{From Boolean Algebra to a Matrix Algebra}
\label{sec:BA2VS}

Matrix Graph Grammars use Boolean algebra mainly because production
actions and graphs are characterized through Boolean matrices and
Boolean operations (\textbf{and} and \textbf{or}). As swaps are linear
transformations, it seems interesting to move to algebra and try to
identify the linear operations (if any). The first thing to do is to
recover productions from swaps.

Recall that a swap can be applied to any host graph
whereas for a production we need to find some elements in the host
graph $(L)$ and guarantee that some others $(K)$ are not present: a
production is a swap together with some restrictions.

\begin{proposition}
  \label{prop:SwapProdEquivalence}
  Let $p = (\mathcal{L}, e + i r) = (L + iK, e + ir)$ be as in
  previous sections a production and $\omega = P(p) = (e + r +
  \mathbf{1}) + i(e + r)$ its associated swap, applied to the host
  graph $G$. The image of the production -- $\mathcal{R} =
  p(\mathcal{L})$ -- is given by:
  \begin{equation*}
    \begin{cases}
      H = p(G) = \sigma.\,\omega + G & \mbox{(graph transformation)}
      \\
      QR = (\sigma.K)G = (\sigma.L)(G + \mathbf{1}) = \mathbf{0} &
      \mbox{(application conditions)}
    \end{cases}
  \end{equation*}
\end{proposition}

\noindent \emph{Proof} \\
$\square$According to Prop.~\ref{prop:nonDet} the application can be
deterministic, non-deterministic or impossible, depending on
$\sigma$. In~\cite{MGG_Book}, Chap.~6, it is proved that a direct
derivation can be defined (equiv., a production can be applied) if the
production is well-defined (compatible) and its LHS is found in the
host graph (and the nihilation matrix is not present in the host
graph).

The first application condition $QR = 0$ is compatibility. If we limit
ourselves to nodeless MGGs, this condition will be always
fulfilled. The second condition demands the non-existence of $K$ in
$G$. It is equivalent to $(\sigma.K)(G + \mathbf{1}) = \sigma.K$. The
third condition demands the existence of $L$ in $G$ as it is
equivalent to $(\sigma.L)G = \sigma.L$. \proofend

\noindent \textbf{Remark}.$\square$The \emph{graph transformation}
part is linear while the \emph{application condition} part is
non-linear. Application conditions and their generalizations
(so-called \emph{graph restrictions}) are studied
in~\cite{MGG_Fundamenta}, and also in Chaps.~8~and~9
in~\cite{MGG_Book}. \proofend



The next step is to characterize the application of a sequence with a
finite number of productions, $s = p_n; \ldots; p_1$. There will be
again two parts, one linear with the actions and one non-linear with
the restrictions (see Th.~\ref{th:sequenceApplicability} below). In
particular we shall guarantee in the propositions that follow that the
sequence is compatible, coherent and that the initial digraph is
contained in $G + i(G + \mathbf{1})$, being as always $G$ the host
graph.

\begin{proposition}
  \label{prop:coherenceInSequences}
  With notation and hypothesis as in Prop.~~\ref{prop:nonDet},
  the same conclusions hold if the operator $C(s) = C^+(s) + i
  C^-(s)$ that assigns one element of $\mathbb{M}^n$ to the sequence
  $s$ is considered:
  \begin{align}
    C^+(s) = 1 + \prod_{j=1}^n &\left( R^\sigma_j
      \underset{j+1}{\overset{n}{\nabla}} \left[ \left( e^\sigma_x +
          \mathbf{1} \right) r^\sigma_y \right] + L^\sigma_j
      \underset{1}{\overset{j-1}{\bigtriangleup}} \left[ \left(
          {r}^\sigma_x + \mathbf{1} \right) e^\sigma_y \right]  +
    \right. \nonumber \\
    \label{eq:35}
    &+ \left. L_j^\sigma R_j^\sigma
      \underset{j+1}{\overset{n}{\nabla}} \left(
        e^\sigma_x + \mathbf{1} \right) r^\sigma_y
      \underset{1}{\overset{j-1}{\bigtriangleup}} \left(
        {r}^\sigma_x + \mathbf{1} \right) e^\sigma_y \right),
  \end{align}
  \begin{align}
    \label{eq:36}
    C^-(s) = \mathbf{1} + \prod_{j=1}^n &\left( Q^\sigma_j
      \underset{j+1}{\overset{n}{\nabla}} \left[ \left(
          r^\sigma_x + \mathbf{1} \right) e^\sigma_y \right] +
      K^\sigma_j \underset{1}{\overset{j-1}{\bigtriangleup}}
      \left[ \left( e^\sigma_x + \mathbf{1} \right) r^\sigma_y \right]
      + \right. \nonumber \\
  &+ \left. K^\sigma_j Q^\sigma_j
      \underset{j+1}{\overset{n}{\nabla}} \left(
          r^\sigma_x + \mathbf{1} \right) e^\sigma_y
        \underset{1}{\overset{j-1}{\bigtriangleup}} \left( e^\sigma_x
          + \mathbf{1} \right) r^\sigma_y \right),
  \end{align}
  being
  \begin{align*}
    \underset{t_0}{\overset{t_1}{\bigtriangleup}} F(x,y) &= \mathbf{1} +
    \prod_{y=t_0}^{t_1} \left( \mathbf{1} + \prod_{x=y}^{t_1} F(x,y)
    \right), \\
    \underset{t_0}{\overset{t_1}{\nabla}} F(x,y) &= \mathbf{1} +
    \prod_{y=t_0}^{t_1} \left( \mathbf{1} + \prod_{x=t_0}^{y} F(x,y)
    \right).
  \end{align*}
\end{proposition}

\noindent \emph{Proof} \\
$\square$ All we have to do to apply Prop.~\ref{prop:nonDet} is to
prove the equivalence between
equations~(\ref{eq:30})~and~(\ref{eq:34}) and
equations~(\ref{eq:35})~and~(\ref{eq:36}), respectively. Also, we have
to check sameness of $\bigtriangleup'$ and $\nabla'$ in
Prop.~\ref{prop:nonDet} with $\bigtriangleup$ and $\nabla$ in this
Proposition. To this end, simply use the following identities from
propositional logics:
\begin{equation*}
  x_j \wedge x_k = x_j \cdot x_k, \qquad \bigvee_{j=1}^n x_j = 1 +
  \prod_{j = 1}^n \left( 1 + x_j \right), \qquad \overline{x} = 1 +
  x.
\end{equation*}
Both cases $C^+$ and $C^-$ are almost equal. \proofend

A similar reasoning applies to the calculation of the initial digraph
of a coherent sequence and its compatibility in the following two
propositions. as cmomented above, they will be used to characterize
determinism of a sequence of productions.

\begin{proposition}
  \label{prop:initialDigraph}
  Let $s = p_n; \ldots ; p_1$ be a coherent sequence with notation as
  above. Then, the initial digraph is given by:
  \begin{equation}
    \label{eq:37}
    M(s) = M_C(s) + i M_N(s) = \underset{1}{\overset{n}{\nabla}}
    \left[ \left( r^\sigma_x + \mathbf{1} \right) L^\sigma_y + i
      \left( e^\sigma_x + \mathbf{1} \right) \left( T^\sigma_x +
        \mathbf{1} \right) K^\sigma_y \right].
  \end{equation}
\end{proposition}

\noindent \emph{Proof} \\
$\square$ $\blacksquare$

\begin{proposition}
  \label{prop:compatibility}
  Let $s = p_n; \ldots ; p_1$ be a sequence made up of compatible
  productions. Then, $s$ is compatible if $W(s) = \mathbf{0}$,
  where\footnote{The letter $W$ has been chosen because compatibility
    can be understood as well-formedness of the sequence, i.e. the
    sequence does not define an operation that produces something
    which is not a simple digraph: the space is \emph{closed}.}
  \begin{equation}
    \label{eq:38}
    W(s) = \underset{1}{\overset{n}{\nabla}} \left[ \left( e^\sigma_x
        + \mathbf{1} \right) \left( r^\sigma_x + \mathbf{1} \right)
      M_C(s_x) M_N(s_x) \right].
  \end{equation}
\end{proposition}

\noindent \emph{Proof} \\
$\square$ $\blacksquare$

Previous propositions allow us to characterize determinism -- notice
that determinism extends applicability as introduced
in~\cite{MGG_Book}, Chap.~1 --. According to Th.~6.4.4
in~\cite{MGG_Book} all we need is compatibility and coherence of the
productions plus finding the initial digraph in $G + i(G +
\mathbf{1})$.

\begin{theorem}
  \label{th:sequenceApplicability}
  Let $s = p_n; \ldots; p_1$ be a sequence of productions with
  associated swaps $\omega_j, j = 1, \ldots, n$. The image $H$ of the
  sequence when applied to the host graph $G$ is given by:
  \begin{equation*}
    \begin{cases}
      H = s(G) = \sum_{j = 1}^n \sigma_j.\,\omega_j + G &
      \mbox{(graph transformation)} \\
      W(s) = C(s) = M_N(s) G = M_C(s)(G + \mathbf{1}) = \mathbf{0} &
      \mbox{(application conditions)},
    \end{cases}
  \end{equation*}
  being $M(s)$ the initial digraph, $W(s)$ the compatibility
  conditions and $C(s)$ the coherence conditions.
\end{theorem}

\noindent \emph{Proof} \\
$\square$ $\blacksquare$

If we stick to nodeless MGG then the formulas get simpler as $D_i +
\mathbf{1} = T_i = \mathbf{0}$ and $K_i = r_i, i \in \{1, \ldots,
n\}$. Again, the graph transformation part is linear while the part of
the application conditions is non-linear.




\section{MGG and Turing Machines}
\label{sec:MGGandTM}

In this section the relationship between Matrix Graph Grammars and
Turing machines (TM) is studied. Two standard introductions to TMs
are~\cite{Hopcroft} and~\cite{sipser}. We will see how to simulate a
TM using nodeless MGG. The simulation of TMs using MGG is easy except
that the tape in TMs is unbounded.

According to~\cite{sipser}, a Turing machine can be formally defined
as a finite-state machine with a memory medium. It is the 7-tuple
\begin{equation}
  M = \left( Q, \Gamma, \Sigma, \tau, q_0, q_a, q_r \right),
\end{equation}
where $Q$ is the set of possible states, $\Gamma$ is the set of
\emph{tape symbols} (a special one named \emph{blank symbol} belongs
to $\Gamma$, $\sqcup \in \Gamma$), $\Sigma \subset \Gamma \backslash
\{\sqcup\}$ is the set of \emph{input symbols}, $\tau : Q \times
\Gamma \rightarrow Q \times \Gamma \times \{LS,RS\}$ is the (partial)
\emph{transition function},\footnote{$LS$ stands for \emph{left shift}
  and $RS$ for \emph{right shift}. At times the set $\{LS,RS,Nmov\}$ is
  considered, where $Nmov$ allows the machine to stay in the same
  cell. This variation does not increase the machine's computational
  power.} $q_0$ is the \emph{initial state}, $q_a$ is the accept state
and $q_r$ is the reject state. All sets under consideration are
finite, except for the length of the tape. The blank symbol is the
only one allowed to occur infinitely many times in the tape during
computation.

The set of states $Q$ will be modeled with labeled nodes as well as
the sets of symbols $\Gamma$ and $\Sigma$ and the initial, accepting
and rejecting states $q_0$, $q_a$ and $q_r$. Productions will be used
to model $\tau$. There will be as many productions as rows in the
state table of the TM (see table~\ref{tab:TME} for an example). State
tables are a common means to represent transition functions for TMs.

There are several remarks at hand. First of all, finding a match for a
production can be done efficiently. This is so because each production
has a state node that works as a flag indicating the place(s) where
the production can be applied.

Second, it is straightforward to simulate non-determinism. Election
non-determinism is non-determinism as normally defined in
TMs. Allocation non-determinism happens in TMs when (in a
non-deterministic TM) the same row of the state table can be applied
to two (or more) different paths inside the computation tree.

\begin{figure}[htbp]
  \centering
  \includegraphics[scale = 0.55]{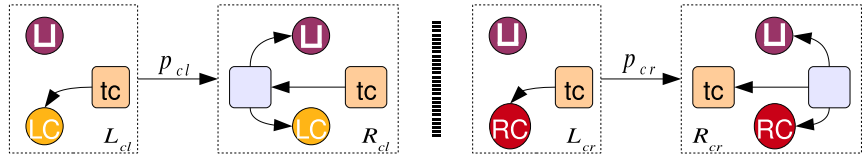}
  \caption{Simulation of an Infinite Tape: Tape Management
    Productions. $p_{cl}$ Adds a Cell to the Left and $p_{cr}$ Adds a
    Cell to the Right}
  \label{fig:tapeProds}
\end{figure}

Third, the tape of any TM has unbounded capacity. On the MGG side, we
want to stick to finite simple digraphs. Two productions (see
Fig.~\ref{fig:tapeProds}) will be responsible for tape enlargement
just in case it is necessary.\footnote{In Fig.~\ref{fig:tapeProds},
  $LC$, $RC$ and $\sqcup$ are labels but $tc$ is not: it just points
  out what nodes are identified by the mapping. For an explanation of
  the meaning of the labels, see the example below.} The idea is to
use the labeled node $LC$ to mark the leftmost cell and $RC$ the
rightmost one. Notice however that nodeless MGG does not allow
addition nor deletion of nodes. 

The point here is that a TM is a \emph{uniform} model of computation
while nodeless MGG is not. There are two possibilities:
\begin{itemize}
\item We just need to add nodes. Deletion of nodes is not necessary
  for TMs. Therefore we may keep the main property of nodeless MGGs
  (no compatibility issues) if we allow the grammar to add nodes to
  the tape.
\item We may use a non-uniform model of computation and have one
  nodeless MGG per number of nodes. The only problem with
  non-uniformity is that various nodeless MGG may have completely
  dissimilar structure. Precisely, this is avoided with productions
  $p_{cl}$ and $p_{cr}$.\footnote{An example of non-uniform model of
    computation is BC. Uniformity conditions have to be imposed to BC
    because otherwise even non-computable functions can be computed by
    small circuits. See~\cite{Vollmer} for the details.}  Observe also
  that addition of nodes have been used just to simulate the infinite
  tape, but not for modeling the behavior of the TM.
\end{itemize}

The rest of the section is devoted to an example that models a copy
subroutine, taken from~\cite{WikiTME}. Its behavior is summarized in
table~\ref{tab:TME}. This TM replicates any sequence of ones inserting
a zero in the middle (the head should be positioned in any of the 1's
that make up the string). For example, it transforms $0000\mathbf{11}0
\longmapsto 0110\mathbf{11}0$.

\begin{table}[htbp]
  \centering
  \begin{tabular}{|c|c|c|c|c|c|c|}
    \hline
    \textbf{ equiv prod } && \textbf{ init state } & \textbf{ tape
      symbol } & \textbf{ print op } & \textbf{ head motion } &
    \textbf{ final state } \\
    \hline
    \hline
    $p_{10}$ && $s_1$ & 0 & Nop  & Nmov & H \\
    \hline
    $p_{11}$ && $s_1$ & 1 & P0 & HL & $s_2$ \\
    \hline
    $p_{20}$ && $s_2$ & 0 & P0 & HL & $s_3$ \\
    \hline
    $p_{21}$ && $s_2$ & 1 & P1 & HL & $s_2$ \\
    \hline
    $p_{30}$ && $s_3$ & 0 & P1 & HR & $s_4$ \\
    \hline
    $p_{31}$ && $s_3$ & 1 & P1 & HL & $s_3$ \\
    \hline
    $p_{40}$ && $s_4$ & 0 & P0 & HR & $s_5$ \\
    \hline
    $p_{41}$ && $s_4$ & 1 & P1 & HR & $s_4$ \\
    \hline
    $p_{50}$ && $s_5$ & 0 & P1 & HL & $s_1$ \\
    \hline
    $p_{51}$ && $s_5$ & 1 & P1 & HR & $s_5$ \\
    \hline
    --      && H    & -- & -- & -- & -- \\
    \hline
  \end{tabular}
  \caption{Copy Subroutine State Table}
  \label{tab:TME}
\end{table}

\begin{figure}[htbp]
  \centering
  \includegraphics[scale = 0.38]{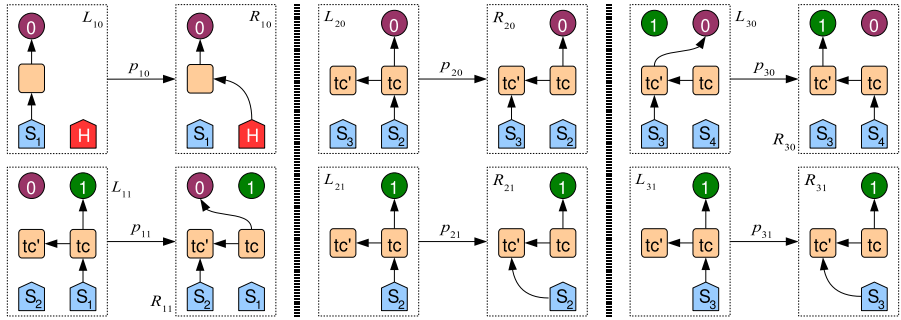}
  \caption{First Six MGG Productions for TM Simulation. Refer to
    Table~\ref{tab:TME} for Their TM Counterparts}
  \label{fig:TMprods1}
\end{figure}

\begin{figure}[htbp]
  \centering
  \includegraphics[scale = 0.43]{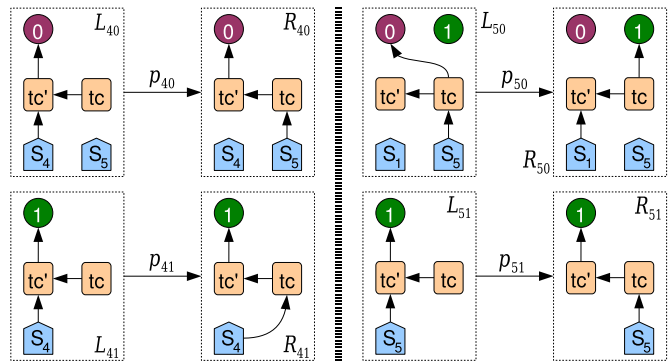}
  \caption{Four More MGG Productions for TM Simulation. Refer to
    Table~\ref{tab:TME} for Their TM Counterparts}
  \label{fig:TMprods2}
\end{figure}

In table~\ref{tab:TME}, the first column is the equivalent MGG
production, which can be found in Figs.~\ref{fig:TMprods1} and
\ref{fig:TMprods2}. The rest of the columns specify the TM
behavior. Column \emph{init state} is the initial state of the TM,
the third column is the tape symbol in the cell being read,
\emph{print op} is the print operation to be carried out in the tape
cell under consideration,\footnote{\emph{Nop} stands for no operation,
  \emph{P0} is ``print zero'' and \emph{P1} is ``print one''.}
\emph{head motion} indicates where the head should move
to\footnote{\emph{Nmov} stands for no movement, \emph{HL} for move
  head left (equiv., move tape right) and \emph{HR} for move head
  right (equiv., move tape left).} and \emph{final state} is the state
assumed by the TM ($H$ is the halting state) which becomes the initial
state of the TM for the next operation. For example, the second row
($p_{11}$) says ``\emph{if your state is $s_1$ and there is a 1 (one)
  in the tape cell, then print a 0 (zero), move to the cell on your
  left and assume state $s_2$}''.

One production simulates each action of the TM. There are
ten, drawn in Figs.~\ref{fig:TMprods1} and~\ref{fig:TMprods2}. A brief
explanation follows:
\begin{itemize}
\item Blue squared nodes represent tape cells. Either a 0 or a 1 have
  to be written on them.  As in this example they can not be blank,
  they should be initialized to 0 (in $p_{cl}$ and $p_{cr}$ on
  Fig.~\ref{fig:tapeProds} we should substitute $\sqcup$ by $0$).
\item Node $S_i$ stands for the $i$th state. It is also used to mark
  which cell the head points to.
\end{itemize}



The set of operations that copies the ones in the word $0\mathbf{11}0$
is $s_c = p_{10}; p_{50}; p_{40}; p_{41};\\ p_{30};p_{31}; p_{cl};
p_{20}; p_{11}; p_{50}; p_{51}; p_{40}; p_{30}; p_{20}; p_{cl};
p_{21}; p_{11}$. This is the sequence that transforms $0\mathbf{11}0$
into $110\mathbf{11}0$. As commented in the introduction, MGG
techniques are available for TMs. For example, it is possible to
calculate $s_c$ -- at least we can guess what productions are
necessary and the number of times that each one appears -- using
\emph{reachability} (see~\cite{MGG_PNGT} or Chap.~10
in~\cite{MGG_Book}).


\section{MGG and Boolean Circuits}
\label{sec:MGGandBC}



In this section we will simulate Boolean circuits (BC) using nodeless
MGG. A standard reference on BCs is~\cite{Vollmer}.

Recall that a BC is a simple digraph. We need only pay attention to
the evolution of the circuit, which is equivalent to encoding the
representation of the Boolean operations permitted in the BC. In
\cite{Vollmer}, the first thing to do when defining a BC is to fix the
Boolean operations allowed. In the present contribution we will
restrict ourselves to the Boolean operations \textbf{and}, \textbf{or}
and \textbf{not}. The same techniques that will be described apply to
any Boolean operation.

There are many ways to encode Boolean operations. We represent the
gates with yellow circles labeled with the corresponding Boolean
operation. The variables are unlabeled circles in blue. Notice that
$x_i$ and $y$ are not labels but a means to visually represent which
nodes on the LHS are mapped to which nodes on the RHS. On the
contrary, $\vee$, $\wedge$, $\neg$, $0$ and $1$ are
labels. 

\begin{figure}[htbp]
  \centering
  \includegraphics[scale = 0.37]{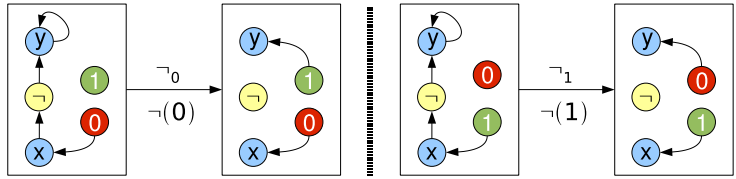}
  \caption{The \textbf{not} Operation}
  \label{fig:notOp}
\end{figure}

Gates labeled with \textbf{not} can only have a single input and a single
output node. Gates of types \textbf{or} and \textbf{and} must have two or more
inputs and a single output. An unlabeled node is an input variable of
some gate if there is an edge from the node to the gate. It will be an
output node if there is an edge from the gate to the node. For
example, $y$ is an output node in $\neg(0)$ in Fig.~\ref{fig:notOp}
and $x$ is an input node.

\begin{figure}[htbp]
  \centering
  \includegraphics[scale = 0.31]{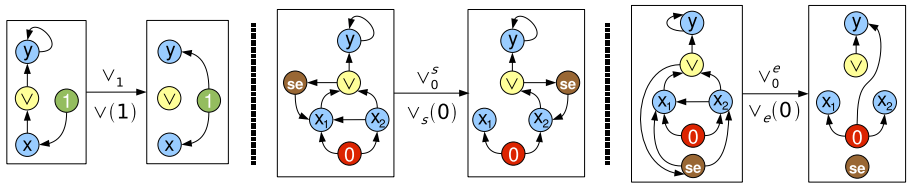}
  \caption{The \textbf{or} Operation}
  \label{fig:orOp}
\end{figure}

\begin{figure}[htbp]
  \centering
  \includegraphics[scale = 0.31]{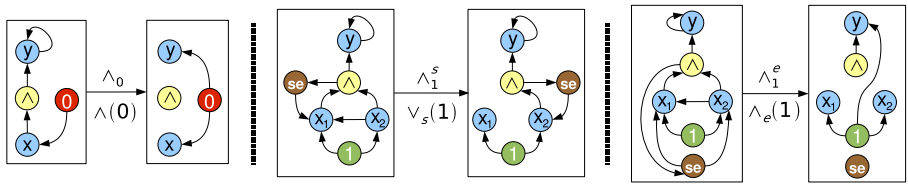}
  \caption{The \textbf{and} Operation}
  \label{fig:andOp}
\end{figure}

Input and output nodes can be in three states: either no value has
been assigned yet (which will be indicated by a self-loop) or it has
value $0$ or value $1$ (represented with an edge from a properly
labeled node). See Figs.~\ref{fig:notOp},~\ref{fig:orOp} and
\ref{fig:andOp} for some examples.

The \textbf{not} operation (see Fig.~\ref{fig:notOp}) is almost
self-explanatory. It is not mandatory to delete the input and output
edges (to the \textbf{not} node) as the rule would not be applicable
to the same elements in the host graph because of the missing
self-loop in node $y$. We do it as a visual aid. The value of the
input variable is kept (the edge is not deleted) because this node
could well be the input of another gate.

We will not limit the number of input nodes to \textbf{or} and
\textbf{and} gates. To ease operations, we will demand an ordering in
their input nodes. One way to achieve this is to introduce a node
\emph{se} (start-end) to mark the first input node and the last input
node for some gate $B$ (= \textbf{and}, \textbf{or}). Every gate of
type \textbf{and} and \textbf{or} will have an edge incident to
\emph{se} (each gate $B$ has its own ordering, but a single node may
belong to the input nodes of several gates). One edge from input node
$x_1$ to input node $x_2$ will mean that $x_1$ goes before $x_2$
considered as an input node for gate $B$. See Figs.~\ref{fig:orOp}
and~\ref{fig:andOp} for some examples.


The \textbf{or} operation is represented in Fig.~\ref{fig:orOp}.
Production $\vee(1)$ forces the result to be $1$ as soon as any input
node has this value. Grammar rule $\vee_s(0)$ decreases the number of
input nodes if both are zero. Notice that $\vee_s(0)$ is applied in
strict order in the input nodes. Rule $\vee_e(0)$ assigns a $0$ to
the output gate when only the starting and the ending input nodes are
left.

As in the \textbf{not} gate, values assigned to the input nodes are
not removed because a single node can be an input for several
gates. The same remark applies to the \textbf{and} operation, which is
described next.

The productions that encode the \textbf{and} operation are depicted in
Fig.~\ref{fig:andOp}. Their interpretation is very similar to that of
the \textbf{or} operation.

The productions that appear in Figs.~\ref{fig:notOp},~\ref{fig:orOp}
and~\ref{fig:andOp} simulate Boolean circuits
non-deterministically. This non-determinism guarantees that the
operations are performed optimally: only those nodes necessary to
evaluate the BC will be calculated and in the precise order.

The complexity of a BC is measured in terms of the size (number of
gates) and the depth (length of the longest directed path). There are
theorems that relate lower bounds on BCs with lower bounds on TMs
(see~\cite{Vollmer}). The non-deterministic encoding of BCs given in
this section should not affect this complexity measures. Nevertheless,
it should be clear how to transform non-determinism into determinism
using some control nodes.

\section{Conclusions and Future Work}
\label{sec:conclusions}

In the present contribution we have introduced MGG as a formal grammar
as well as a model of computation. We have also seen that
non-determinism in MGG can be approached through so-called
\emph{relabeling}. It has also been proved that nodeless MGG is
capable of simulating deterministic and non-deterministic Turing
machines and Boolean Circuits. As a side effect, all MGG techniques
are at our disposal to tackle different problems in Turing machines
and Boolean Circuits. As commented in Sec.~\ref{sec:intro}, we can
(partially) address coherence, congruence, initial state
characterization, sequential independence, reachability and some
others. Something similar but for Petri nets is done
in~\cite{MGG_PNGT} or in~\cite{MGG_Book}, Chap.~10. However, we have not
gone into these topics in depth. The theory developed so far can be
applied straightforwardly in some cases, while in others some further
research is needed.

The use of MGG opens the door to the introduction of analytical and
algebraic techniques over finite fields. We are particularly
interested in representation theory and abstract harmonic
analysis.

Apparently, reachability and the state equation appear to be helpful
concepts to address complexity problems. In~\cite{MGG_PNGT}, the
reason why the state equation is a necessary but not a sufficient
condition is pointed out. It seems natural to move from linear algebra
(state equation) to linear programming (positivity restrictions on
variables).  \\ \

\noindent \textbf{Acknowledgements}. I would like to express my
gratitude to the open source and copyleft communities, in particular
to the people behind the text ``editor'' Emacs
(\url{http://www.gnu.org/software/emacs/}), TeX Live (the Linux \LaTeX
distribution, \url{http://www.tug.org/texlive/}), OpenOffice
(\url{http://www.openoffice.org}) and Linux
(\url{http://www.linux.org/}), especially the Ubuntu distro
(\url{http://www.ubuntu.com/}).


\end{document}